\documentclass[pre,aps,floats,superscriptaddress,floatfix]{revtex4}
\usepackage{amssymb,amsmath}
\usepackage{amsmath,amssymb}
\usepackage{graphicx}
\usepackage{psfrag}
\usepackage{color}
\usepackage{dcolumn}
\usepackage{bm}
\usepackage[normalem]{ulem}

\def\beq{\begin{equation}}
\def\eeq{\end{equation}}
\def\bea{\begin{eqnarray}}
\def\eea{\end{eqnarray}}
\begin{document}
\title{Pinned or moving: states of a single shock in a ring}
\author{Parna Roy}\email{parna.roy14@gmail.com}
\affiliation{Condensed Matter Physics Division, Saha Institute of
Nuclear Physics, Calcutta 700064, West Bengal, India}
\author{Anjan Kumar Chandra}\email{anjanphys@gmail.com}
\affiliation{Malda College, Malda 732101, West Bengal, India}
\author{Abhik Basu}\email{abhik.123@gmail.com}
\affiliation{Condensed Matter Physics Division, Saha Institute of
Nuclear Physics, Calcutta 700064, West Bengal, India}

\date{\today}
\begin{abstract}
 Totally asymmetric exclusion processes (TASEP) with open boundaries are known to exhibit 
moving shocks or delocalised
domain walls (DDW) for sufficiently small equal injection and extraction rates. 
In contrast
 TASEPs in an inhomogeneous ring have been shown to display pinned shocks or localised domain walls 
(LDW) under similar conditions [see, e.g., H. Hinsch and E. Frey, {\em Phys. 
Rev. Lett.} 
{\bf 97}, 095701 (2006)]. By studying periodic exclusion processes composed of 
a 
driven (TASEP) and a diffusive segments, we uncover smooth transitions between 
LDW and DDW; the latter mimics DDWs in an open TASEP, controlled essentially by the 
fluctuations in the diffusive segment. Mean-field theory 
together with Monte Carlo simulations  are employed to characterize the 
emerging nonequilibrium steady states. Our studies provide an explicit route to 
control the degree of shock fluctuations in periodic systems, and should be 
relevant in cell biological transport where the availability of molecular 
motors is the rate limiting constraint.
\end{abstract}
\maketitle

\section{Introduction}

Totally asymmetric simple exclusion process (TASEP) with open boundaries 
 was originally proposed as a simple model for the motion of 
molecular motors in eukaryotic cells~\cite{pipkin}. Subsequently, it was 
reinvented as a paradigmatic one-dimensional model for nonequilibrium 
statistical mechanics, that shows boundary induced phase 
transitions characterised by $\alpha$ and $\beta$, the entry 
and exit rates~\cite{derrida}. Extensive Monte-Carlo simulations (MCS) 
supplemented by mean field theories (MFT) reveal that for $\alpha=\beta<1/2$, 
the steady state density profile shows a {\em  moving shock} or {\em delocalised domain wall} (DDW) that 
moves randomly and unrestrictedly along the whole TASEP~\cite{tasep-basic}. 
This is usually explained in terms of completely uncorrelated entry and exit events.

Different situations can emerge in TASEPs in closed rings. For TASEP on a 
homogeneous ring, translational invariance ensures a uniform
mean density along the ring. Macroscopically nonuniform steady state densities 
can emerge only when
translational invariance is explicitly broken. This can happen in a variety of 
ways. For instance, TASEP in a ring with a single bottleneck or a defect site with a 
lower hopping rate than the remaining system can show a {\em pinned shock} or {\em localised domain 
wall} (LDW) for moderate average densities. In yet another manifestation of breakdown of translational 
invariance, a closed system composed of two segments of equal size -  a TASEP 
and a diffusive lane with exclusion, executing what is known as {\em symmetric 
exclusion process} or SEP - also shows an LDW in the steady states for moderate 
average densities~\cite{hinsch}. In contrast, inhomogeneous TASEP in a ring with 
strict particle number conservation can  display two or more 
DDWs only when there are more than one bottlenecks of equal 
strength~\cite{niladri1,tirtha1}.  
To 
our knowledge, a single DDW in a closed heterogeneous 
TASEP has never been observed
 till the date.
 It is thus
pertinent to ask: can a single LDW observed in closed TASEPs in 
heterogeneous rings be converted into a {\em single} DDW, resembling 
open TASEPs? If so, under what situations and how - is it a {\em smooth} or {\em sudden} transition?

In the present work, we address this issue systematically by studying a class of conceptual models constructed by us. 
We principally focus on the nature of a single shock in a 
periodic 
system with strict number conservation and study inter-conversions between an 
LDW and a DDW. To this end, we propose and construct three different but 
related periodic model, each 
composed of a TASEP part and a diffusive part and 
study the ensuing nonequilibrium steady states (NESS). All of these models fall in the class of a TASEP 
connected to a reservoir with finite resources (i.e., a given total number of 
particles), with the reservoir having its own internal dynamics modeled as one-dimensional (1D) 
diffusion. We show that in all the models studied here (i) a single DDW can be formed 
in a 
periodic system with large  {\em average particle content} in the 
diffusive segment; the latter can be used to control interconversion of LDW and 
DDW, (ii) this interconversion between LDW and DDW is {\em continuous}, i.e., 
the span of a DDW can be continuously shrunk to zero converting  it into an LDW. All the  models that we construct and study here bear out this physical picture, underscoring 
the robustness of the mechanism for deconfinement of an LDW elucidated in this article. 
The remaining part of this article is structured as follows: In 
Sec. \ref{models} we discuss construction of our models, which are (i) Model IA and Model IB, where the dynamics in the
diffusive segment does not respect any exclusion principle, and (ii) Model II, which enforces exclusion in the diffusive
channel.  In Sec. \ref{density1} the steady state density profiles of Model I is discussed and in 
Sec. \ref{density2} the steady state density profiles of Model II is discussed. We summarise and discuss our results in Sec.~\ref{summary}. Some 
calculational details including phase diagrams are given in Appendices at the end.

\section{Models for periodic exclusion processes}\label{models}

We consider a periodic 1D model  composed of a driven 
 TASEP (${\cal T}$) and a diffusive  (${\cal S}$) parts. In ${\cal T}$, 
hopping of particle  is unidirectional at unit rate that is subject to 
exclusion, where as in $\cal S$  particles 
jump independently and randomly to the neighbouring site at rate $D$ with 
equal probability to left and 
right. Particle dynamics in $\cal S$ may or may not be subject to exclusion. We 
consider both the possibilities for $\cal S$. Particle exchanges between $\cal T$ and $\cal S$ are allowed at the junction with dynamical rules governed by 
the 
originating site. Depending upon the detailed dynamical rules for $\cal S$ we 
consider three different models as described below. 

\subsection{Domain walls in closed heterogeneous TASEPs}


Before we embark upon our studies, it is useful to recall the existing studies and results on domain walls in closed inhomogeneous TASEP.
Some of these model studies have a single heterogeneity or a bottleneck in the form of a point~\cite{lebo} or extended~\cite{gautam} bottleneck, or an intervening diffusive segment~\cite{hinsch}. All these display a single LDW for appropriate choices of the model parameters and particle number densities. Other studies include multiple bottlenecks, point~\cite{niladri1} or extended~\cite{tirtha1}, and can display not just an LDW, but multiple DDWs as well in some regions of the phase space. The DDWs in these models always appear more than one in number. These contrasting behaviours are by now well-understood within MFT. For instance, in the models of Refs.~\cite{hinsch,lebo,gautam}, where only a single LDW may be observed under appropriate conditions, strict particle number conservation ensures  a unique solution for the domain wall position within the MFT that is consistent with an LDW. In 
contrast, for models in Refs.~\cite{niladri1,tirtha1}, where multiple DDWs can be found for specific choices of the parameters, particle number conservation does not yield unique solutions for the individual domain wall positions, but rather gives only relations between them leaving them overall undetermined. Thus there would be many solutions of the positions of the domain walls, all of which maintain the strict particle number conservation. Since all these solutions are equally probable and visited by the systems in course of time due to the inherent stochasticity of the underlying dynamics, long time averages yield envelops of these all possible domain wall solutions which are nothing but DDWs. Thus these DDWs necessarily appear more than one in number. All these are in contrast to an open TASEP, where a single system-spanning DDW and not an LDW can be observed. The latter is ascribed to the lack of strict particle number and uncorrelated entry and exit events in open TASEP. It is thus tempting to 
conclude that in a closed heterogeneous TASEP there cannot be 
a transition between a conventional single LDW to a single DDW that resembles a single DDW in an open TASEP. We however argue below that the existing
arguments that favour a single LDW in a closed heterogeneous TASEP do not consider the particle number fluctuations in the TASEP segment of the 
heterogeneous ring (which is non-zero even though the total particle number remains conserved). When the latter is accounted for, possibilities of 
delocalisation of an LDW opens up. Indeed for sufficiently large fluctuations in the TASEP segment - that is controlled by the fluctuations in the non-TASEP part of the ring due to overall particle number conservation -  an LDW fully delocalise to take the form of a single DDW that spans the whole length of the TASEP segment. In order to establish this behaviour, we propose and study a class of conceptual models below.

\subsection{Models}\label{models}

\begin{figure}[!htbp]
 \includegraphics[width=18cm,height=13.0cm,angle=0]{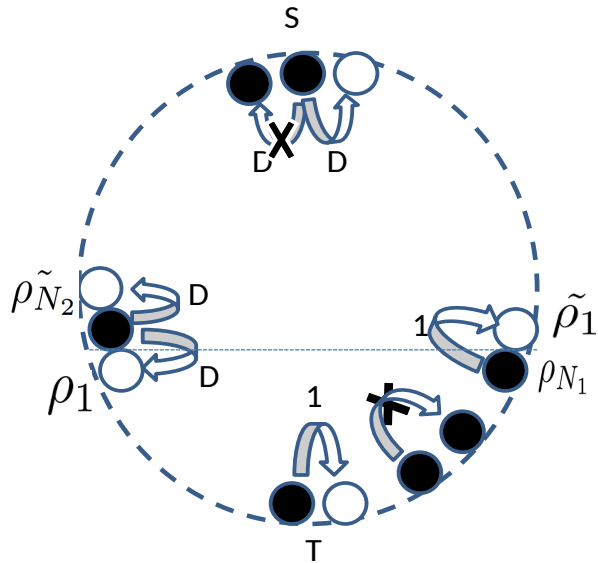}
 \vskip-4cm
  \caption{Schematic diagram of the  models.  Particles in segment $\cal S$ diffuse, where as in $\cal T$ they undergo  dynamics controlled by asymmetric exclusion processes. }
  \label{diag1}
\end{figure}

We study two related conceptual models, each of which consists a diffusive part $\cal S$ 
and a driven part $\cal T$; see Fig.~\ref{diag1} for a generic schematic 
diagram for all the two models we consider here.

The asymmetric segment $\cal T$ has $N$ sites, where as the diffusive segment $\cal S$ has $\tilde N=rN$ sites with 
$r$ can be smaller or larger than unity. In 
order to define the models formally, we denote the location of the lattice sites 
and occupation numbers 
by $x_i\in [1,N]$ and $n_i$ for $\cal T$ and $\tilde{x_i}\in [1,rN],\,r\geq 
1$ and $\tilde{n_i}$ for $\cal S$ respectively. Total number of particles 
\begin{equation}
N_p=\sum_{i=1}^{N_1}n_i+\sum_{i=1}^{N_2}\tilde{n_i}
\end{equation}
is conserved.  

The driven segment $\cal T$ is identical in both the models. These two 
models are however 
distinguished by their respective diffusive segments $\cal S$. In details:

(i) In Model I,  Unlike in $\cal T$, we do not impose
any condition of exclusion in 
$\cal S$, 
i.e., a site in $\cal S$ is here allowed to accommodate any number of particles 
without restrictions. 
As a result,   
the over all particle density $n_p=\frac{N_p}{N(1+r)}$ is {\em not} 
restricted to $[0,1]$. Furthermore, particles can exit $\cal T$ at a given rate $\beta$ to enter into $\cal S$ that is completely unaffected by the occupation of the site 
$\tilde i =1$ in $\cal S$. Thus, $\beta$ is a tuning parameter for Model I. Thus, the NESS of Model I are parametrised by three parameters: 
$D, n_p$ and 
$\beta$.


(ii) In contrast in Model II, we impose exclusion in $\cal S$. Thus 
the dynamics in the whole 
ring in Model II is subject to exclusion -  
each site in $\cal T$ or $\cal S$ can be occupied by at most one particle.
Hence, the particle density $n_p=\frac{N_p}{N(1+r)} \in [0,1]$ necessarily, 
since there could be at most one particle per site. This directly 
generalises the model studied in Ref.~\cite{hinsch}. Notice that the rate at which particles can exit $\cal T$ and move to $\cal S$ 
{\em does depend} on the 
occupation at $\tilde i =1$ - a direct consequence of exclusion in $\cal S$. Thus Model II is a two-parameter model - $n_p$ and $D$.  For reasons similar to 
those discussed in Ref.~\cite{hinsch}, we let diffusivity $D$ scales with the 
system size that ensures a non-zero steady state current in the system (see 
also below).

All the models are shown to display similar deconfinement of LDWs as the 
{\em total
particle content} of $\cal S$ rises, a feature that is attributed to enhanced 
fluctuations in $\cal S$ (see below).

\section{Steady state density profiles}

We use MFT together with extensive Monte-Carlo simulation (MCS) of our model to
obtain the steady state density profiles. In MFT approaches,
the system is considered as a collection of one TASEP ($\cal T$) with open boundaries having effective 
entry and
exit rates and a diffusive lane ($\cal S$) again with open boundaries
 having its own effective entry and exit rates~\cite{hinsch}. These effective 
rates are determined
from the condition of particle current conservation  in the steady states. We then
 use them in conjunction with the known results
for TASEP and diffusive lane with open boundaries to obtain
the density profiles here. An isolated open TASEP in
steady state can be in three different states, the low density (LD), high density (HD) and maximal current (MC)
phases; we expect to find analogues of these phases for $\cal T$. 

We denote the steady state density at a particular site 
$m$ in $\cal T$   as $\rho_m = \langle n_m\rangle$ and in $\cal S$ as
$\tilde\rho_m=\langle \tilde n_m\rangle$, where $\langle..\rangle$ represent 
time-averages in the NESS. MF analysis entails taking continuum limit with 
$\rho(x)$ and $\tilde \rho (\tilde x)$ as the densities in $\cal T$ and $\cal 
S$, where $x=i/N$ and $\tilde x=\tilde i/N$. In the thermodynamic limit (TL), $L\gg 1$ 
and consequently $x$ and $\tilde x$ vary effectively continuously with $0\leq x\leq 1,\,0\leq \tilde x \leq r$. We further
introduce the following notations for the stationary  densities at 
the junctions B and A respectively; $\rho (x=0)=\alpha$ and $\rho (x=1)=(1-\beta)$ at $\cal T$ 
according 
to the standard TASEP convention and $\tilde{\rho} (\tilde x=0)=\gamma$ and  
$\tilde{\rho}({\tilde x= r})=\delta$ for $\cal S$. 
In what follows 
below, we use the continuum labeling $x$ for the densities in NESS.  The MFT
analysis is complemented by extensive MCS studies using random sequential 
updates.

\subsection{Domain walls in Model I}\label{density1}


In Model I, multiple occupancy in each site of $\cal S$ is allowed. In NESS, $\tilde \rho(\tilde x)$ is given by a linear profile in MFT:
\begin{equation}
 \tilde{\rho}(\tilde x)=\delta+(\gamma-\delta)\tilde x/r.\label{sepdengen}
\end{equation}
The corresponding current in $\cal S$ is given by
\begin{equation}
 J_{\cal S}=(\gamma-\delta)D/Nr.
 \label{diffcur}
\end{equation} 
 This must be equal to $J_{\cal T}\sim {\cal O} (1)$ in 
NESS. Now $J_{\cal S}$, as given in (\ref{diffcur}),  can be ${\cal O} (1)$, e.g., when  (i) $\gamma,\;\delta\sim {\cal O}(N)$, so that the difference 
$\gamma -\delta \sim {\cal O}(N)$ together with $D\sim {\cal O} (1)$. We call this Model IA or (ii) $D\sim 
{\cal O}(N)$, but $\gamma,\;\delta\sim {\cal O}(1)$. We call this Model IB. This is 
significantly different from Model IA~\cite{foot11}. 

It  is instructive to first consider the phases and the ensuing phase diagrams of Model IA and Model IB in qualitative terms. Comparing with an open 
TASEP, LD and HD phases are to be found for $\alpha < \beta,\;\alpha <1/2$ and $\beta<\alpha,\;\beta <1/2$, respectively, where as, for 
$\alpha,\;\beta \geq 1/2$ MC phase ensues. Further, with $\alpha=\beta \le 1/2$ one DDW that spans the whole of an open TASEP is found. 
Unlike in an open 
TASEP, where both $\alpha$ and $\beta$ are free parameters that can be tuned, in Model IA or IB $\alpha$ is to be determined from the 
various conditions 
available (see below), while $\beta$ remains free, the other free parameters being $n_p$ and $D$. Consider the phases for $\beta <1/2$ in Model IA. 
By tuning $n_p, D$ $\alpha$ may be varied, as shown below. Thus in the $n_p-D$ plane, regions with $\alpha <\beta < 1/2$ correspond to LD phase; 
the remaining regions where $\alpha >\beta <1/2$ correspond to HD phase. However, there is no MC phase for $\beta <1/2$. Further, regions correspond to $\alpha = \beta<1/2$ imply domain walls, investigation of whose nature is a primary goal of this work. In 
contrast when $\beta > 1/2$, in the $n_p-D$ plane one obtains LD phase in the region with $\alpha <1/2$. In the remaining region for which $\alpha >1/2,\beta >1/2$, MC phase is found. Thus, there are LD and MC phase with no HD phase. This physical picture remains unchanged in 
Model IB. 
In the main text, we focus on the nature of the domain wall and set $\beta <1/2$. The detailed phase diagrams for Model IA and Model IB obtained from MFT and MCS studies are given in the Appendices. 

We assume a domain wall with a mean position $x_w$. Thus steady state density in $\cal T$
\begin{equation}
\rho(x)= \beta + \Theta(x-x_w)(1 -2\beta),\label{tasep-theta}
\end{equation}
where we have used $\alpha = \beta$, which is known since 
 $\beta$ is a given external tuning parameter; here $\Theta(x)$ is the Heaviside $\Theta$-function at $x$. The corresponding steady state current in $\cal T$ is given by
 \begin{equation}
  J_{\cal T}=\alpha (1-\alpha)=\beta(1-\beta). \label{tasep-curr-dw}
 \end{equation}
 The incoming current to site 
$x=0$ at $\cal T$ is 
$\delta D (1-\alpha)$. By using the
current conservation in the steady state, we obtain
\begin{equation}
 \delta D (1-\alpha)=\beta(1-\beta).\label{junca-curr}
\end{equation}
 Since multiple occupancy is allowed in $\cal S$, 
the mean particle number $N_s$ in $\cal S$ in the NESS can be both larger or 
smaller than $\tilde N=Nr$. 
Again due to possible multiple occupancy, both $\gamma$ and 
$\delta$ can be $ {\cal 
O}(N)$. As long as there is a particle in the last site $i_{N}$ of $\cal T$, it 
jumps to first site 
$\tilde{i}_{1}$ of $\cal S$ with probability $\beta$, independent of how many 
particles are already there in  $\tilde{i}_{1}$. Thus, 
$\beta$ is independent of $\gamma$.

\subsubsection{Domain walls in Model IA}
 
\begin{figure}[!htbp]
 \includegraphics[width=10.5cm,height=7.0cm,angle=0]{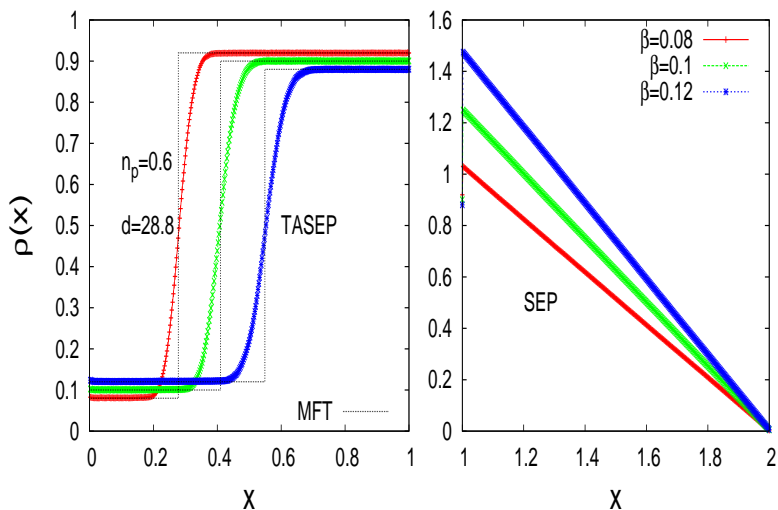}
  \caption{Plot of density profiles with DWs of the active part of 
model IA for $N=400$, $r=1$ and $n_p=0.6$ for different 
$\beta$ and $d=28.8$. These show LDW.}
  \label{LDW2}
\end{figure}

\begin{figure}[!htbp]
 \includegraphics[width=10.5cm,height=7.0cm,angle=0]{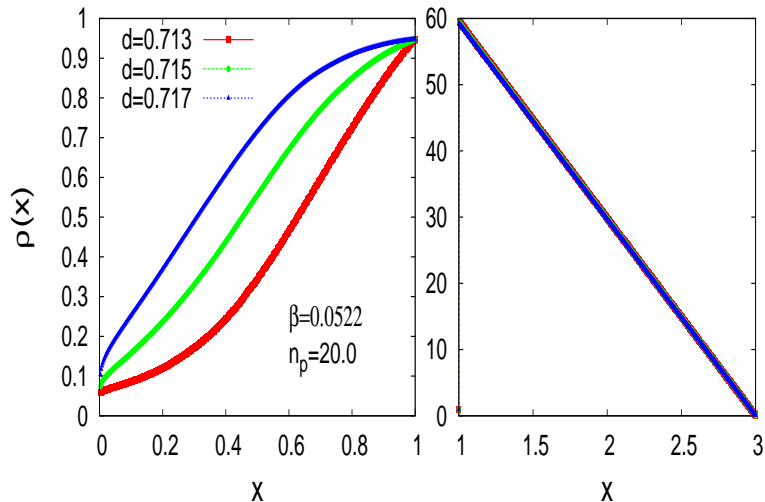}
  \caption{Plot of density profiles with DWs of the active part of 
model IA for $N=400$, $r=2$ and $n_p=20.0$ for different $\beta$ 
and $d=0.0522$ from MCS and MFT studied. MCS studies show DDW, whereas MFT studies predict LDWs. 
}
  \label{DDW2}
\end{figure}

\begin{figure}[!htbp]
 \includegraphics[width=10.5cm,height=7.0cm,angle=0]{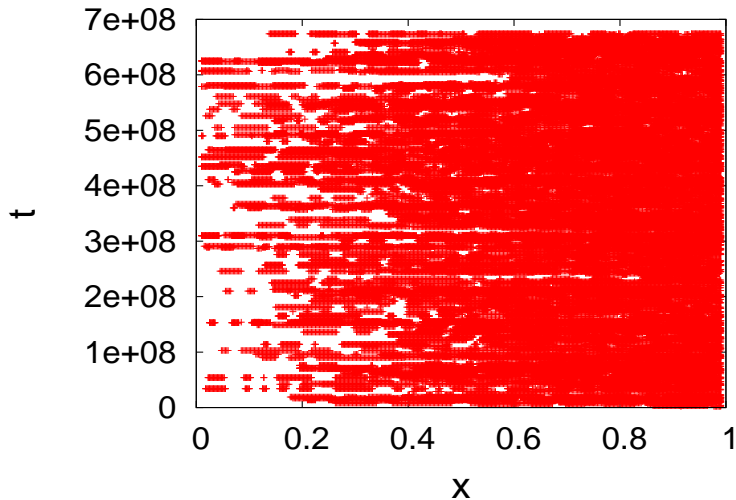}
  \caption{Kymograph for Model IA for $N=400$, $r=2$, $d=0.715$, $\beta=0.0052$ and $n_p=20.0$, obtained from MCS studies. }
  \label{system_2b}
\end{figure}

 Consider a domain wall in Model IA.  Here, $D\sim {\cal O}(1)$. 
  Domain wall position $x_w$ can be calculated from the conservation of total particle number. This requires solving for 
$\tilde \rho(\tilde x)$, which in turn requires knowledge of $\gamma$ and $\delta$ in terms of the model parameters. This can be done by 
using the conservation of the current in the system in the steady state, i.e., by equating $J_{\cal S}$ in (\ref{diffcur}) with  $J_{\cal T}$ given by 
(\ref{tasep-curr-dw}). Straight forward algebra, whose details interested readers can find in Appendix \ref{dw1a}, gives for $x_w$ 
\begin{equation}
 x_w(2\beta-1)+(1-\beta)+\frac{Nr^2}{2D}[\beta(1-\beta)+\frac{\beta}{Nr}]+\frac{\beta r}{2D}=n_p(1+r).
 \label{scdw2}
\end{equation}
Equation~(\ref{scdw2}) clearly yields $x_w$ {\em uniquely} in terms of $\beta, n_p, r, D$.
Since $0<x_w<1$, $n_p$ must be small enough to make (\ref{scdw2}) valid.
Since the mean position $x_w$ is fixed, this implies an LDW within our MFT.
In Fig.~(\ref{LDW2}) we have plotted the stationary density for $N=400$, $r=1$  and $n_p=0.6$, from both MFT and MCS, both of
which clearly show an LDW, which are in good agreement with each other. 
The picture is dramatically different in
Fig. (\ref{DDW2}), where we have plotted the stationary 
density in $\cal T$ for 
$N=400$, $r=1$ and $n_p=20.0$ from MCS studies.
The MCS study unexpectedly shows a DDW, where as MFT results continue to imply an LDW. 
The delocalised nature of the domain wall in MCS study can be confirmed from the corresponding kymograph (Fig.\ref{system_2b}) that clearly 
shows that the spatial extent of the domain wall movements due to its fluctuations spans the entire length of $\cal T$. This is indeed surprising, since all previous studies on closed inhomogeneous TASEP reported that when there is {\em only one} imhomogeneity - point or extended - at most one domain wall in the form of an LDW can be observed. Before 
we analyse and explain this paradoxical behaviour, below we first consider Model IB to check the robustness of a single DDW.

\subsubsection{Model IB}

\begin{figure}[!htbp]
 \includegraphics[width=10.5cm,height=7.0cm,angle=0]{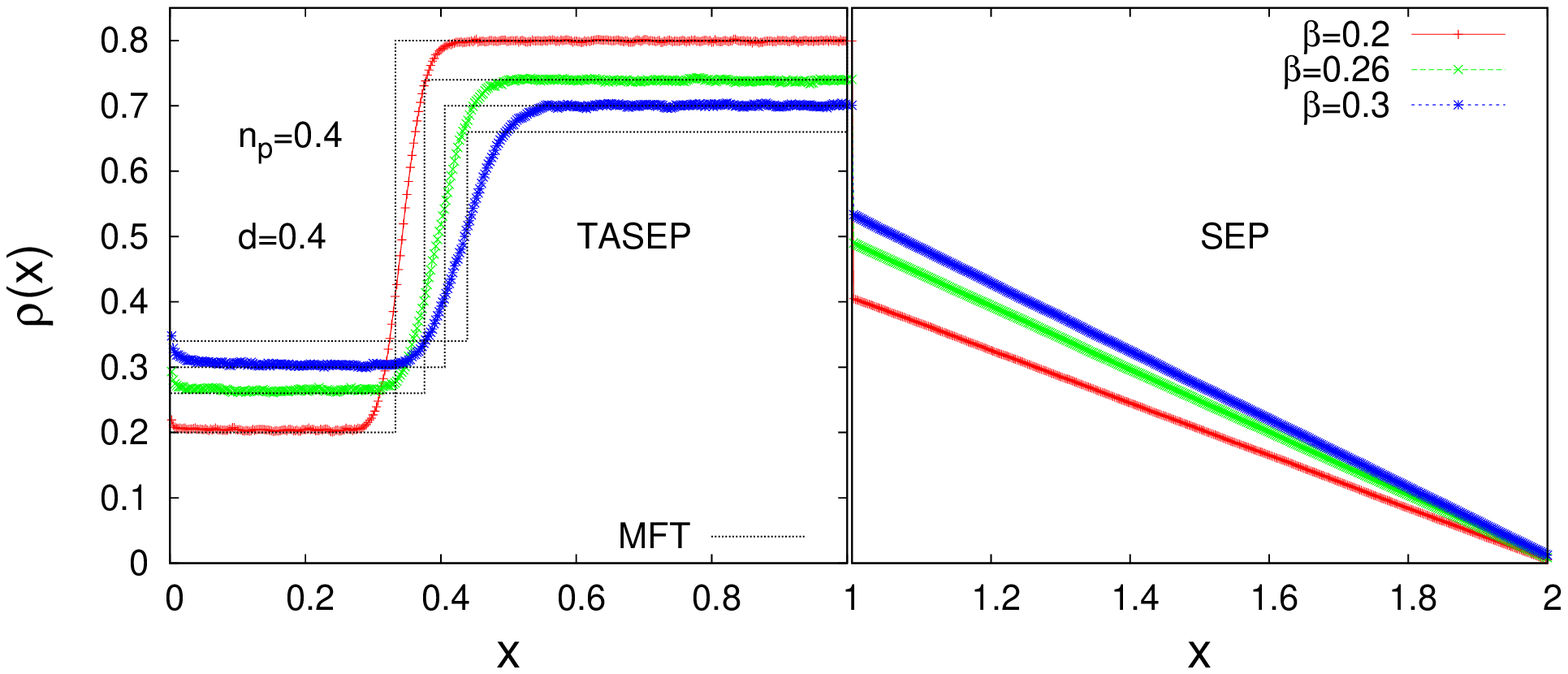}
  \caption{Plot of density profiles for the DW in $\cal T$ in Model IB 
 for $N=400$, $r=1$ and $n_p=0.4$ for different $\beta$ and 
$d=0.4$, from both MCS and MFT studies. Both studies reveal LDWs.}
    \label{LDW1}
\end{figure}

 Let us now study Model IB, when $D$ scales with system size 
$N$; we set $D=d N$ with $d\sim {\cal O} (1)$. Following the logic outlined for Model IA above and using the overall particle number conservation, we obtain
\begin{equation}
 x_w(2\beta-1)+(1-\beta)+\frac{r^2}{2d}
 [\beta(1-\beta)]=n_p(1+r).
 \label{dw2}
\end{equation}
Since $\beta$ is a fixed model parameter, $x_w$ is uniquely determined, implying 
an LDW. In Fig.~\ref{LDW1}, we have plotted $\rho(x)$ versus $x$  for $N=400$, $r=1$ and $n_p=0.4$ for different $\beta$ and $d=0.4$, from both MCS 
and MFT studies. Unsurprisingly, we find 
an LDW; MFT and MCS results agree with each other well. 

\begin{figure}[!htbp]
 \includegraphics[width=10.5cm,height=7.0cm,angle=0]{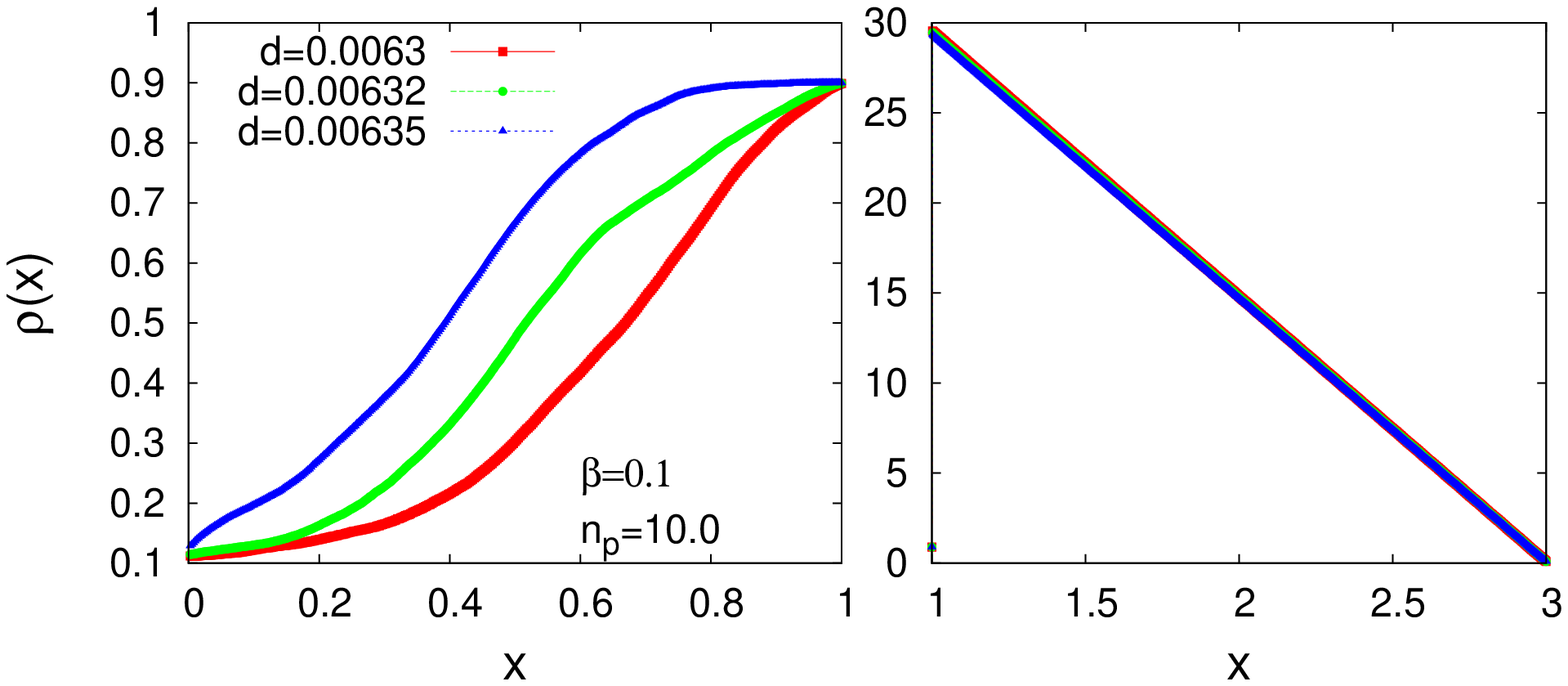}
  \caption{Plot of density profiles for DW in $\cal T$ of Model 
IB for $N=400$, $r=2$ and $n_p=10.0$ for different $d$ and $\beta=0.1$.
  MCS studies exhibit  DDWs. In contrast, MFT yields LDWs.}
  \label{DDW1}
\end{figure}

\begin{figure}[!htbp]
 \includegraphics[width=10.5cm,height=7.0cm,angle=0]{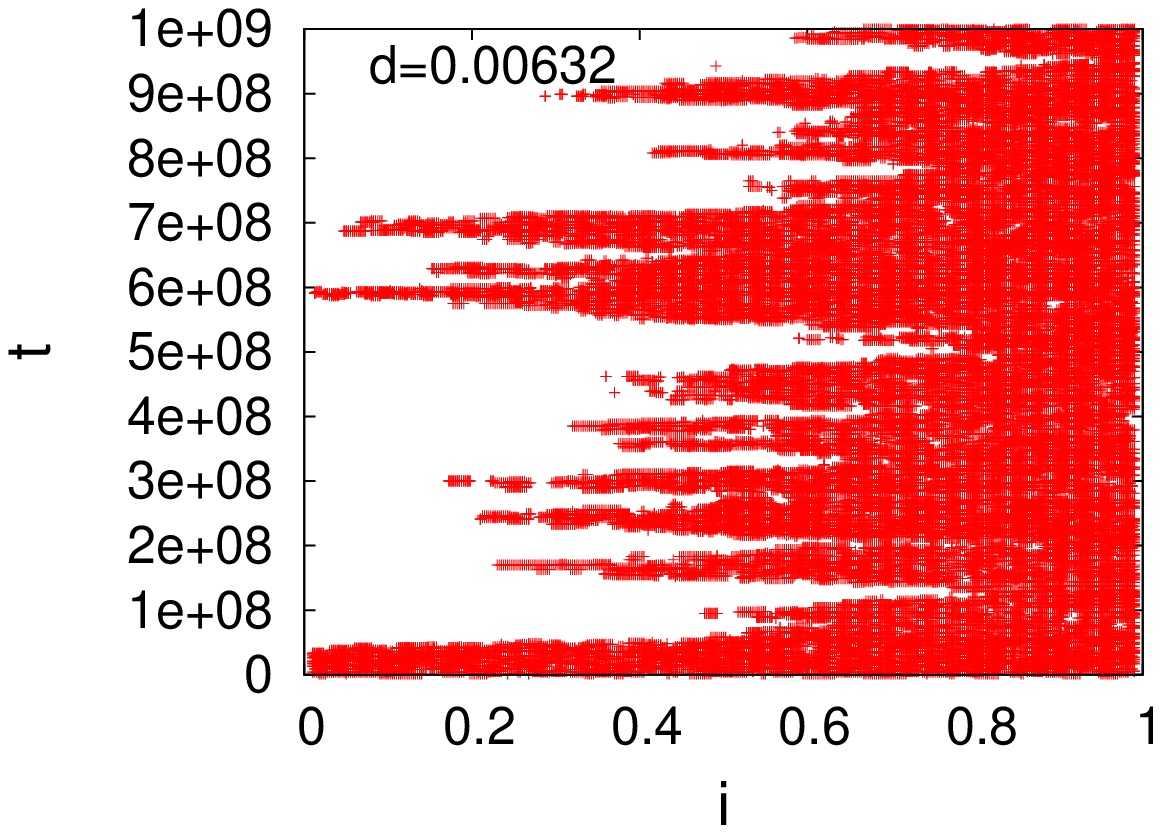}
  \caption{Kymograph for Model IB for $N=400$, $r=2$, $d=0.00632$, $\beta=0.1$ and $n_p=10.0$, obtained from MCS studies. }
  \label{system_2}
\end{figure}

Consider now Fig.~\ref{DDW1}, where we have plotted $\rho(x)$ versus $x$ for $N=400$, $r=2$ and $n_p=10.0$ for different $d$ and $\beta=0.1$. Unexpectedly, we obtain a single DDW from the MCS study, where as the MFT result still predicts an LDW. That $\rho(x)$ in Fig.~\ref{DDW1} is indeed a DDW can be confirmed from the corresponding kymograph in Fig.~\ref{system_2} that clearly shows that the fluctuation of the domain spans the entire length of $\cal T$, confirming a DDW. Therefore, 
the existence of a single DDW, being independent of the precise dynamics in $\cal S$, is fairly robust. 


\subsubsection{Pinned  or moving shocks?}\label{ldwddw}

This conundrum between LDW and DDW can be explained when fluctuations are taken into account. We discuss this in details now. Strict particle number 
conservation in Model IA and IB ensures unique determination of $x_w$ that must imply an LDW. On the other hand, in an open TASEP, total particle number is conserved only on ``average''. Since for any $x_w$ between 0 and 1 
$J_{\cal T}$ is same, the domain wall position can move anywhere in the system, keeping the current in the TASEP unchanged. Furthermore, 
since the domain wall position fluctuations cover the entire length $N$ of the TASEP, the size of the associated number fluctuations is ${\cal O}(N)$. In order for a single DDW to exist in $\cal T$, fluctuations of $x_w$ is to occur over a scale comparable to $N$, which in turn means
 fluctuating particles in $\cal T$. Since total number 
of particles is constant, fluctuations in $\cal T$ is bounded by 
fluctuations in the mean particle number $N_s$ in $\cal S$ with $N_s$ given by
\begin{equation}
 N_s=\int_0^r \, \tilde \rho(x)\,d\tilde x= \int_0^r\,[\delta + (\gamma - \delta)\tilde x/r]\,d\tilde x.
\end{equation}
With values of $\gamma$ and $\delta$ already obtained in MFT above, $N_s$ can be obtained for Model IA and Model IB. For example in Model 
IA $N_s$ is given by 
\begin{equation}
N_s=N\left[\frac{Nr^2}{2D}[\alpha(1-\alpha)+\frac{\alpha}{Nr}]+\frac{r\alpha}{2D}\right].\label{sep-tot1}
\end{equation}
For Model IB since $D$ scales with system size in the TL $D\to \infty$. Total particle in $\cal S$, $N_s$ is given by 
\begin{equation}
N_s=N\left[\frac{Nr^2}{2D}\alpha(1-\alpha)\right]=N\left[\frac{r^2}{2d}\alpha(1-\alpha)\right].\label{sep-tot1}
\end{equation}
Thus, for large $N$ and large $r$, $N_s\sim N^2 r^2$ in Model IA, and $N_s\sim Nr^2$ in Model IB. 
Now, for a DDW to exist in $\cal T$ of a span of size ${\cal O}(N)$, the associated particle number 
fluctuations in $\cal T$ are $\Delta\sim {\cal O}(N)$. In order to maintain 
overall particle number conservation, $\cal S$ must also have particle number 
fluctuations of size $\Delta\sim {\cal O}(N_s)$. That the variance of the particle number in $\cal T$ rises monotonically with the variance 
of the particle number in ${\cal S}$ in the steady states is easily seen in Fig.~\ref{var1a} and Fig.~\ref{var1b}, respectively, for Model IA 
and Model IB. In fact, within our numerical accuracy, these two are found to be equal. 
Assuming non-interacting particles, fluctuations $\Delta$ in the total number 
of particles in $\cal S$ should scale with $\Delta\sim \sqrt{N_s}$. 
Clearly, in the limit $ r<<1$, $\Delta<<1$, fluctuations in $x_w$ is 
negligible; hence an LDW is observed. In contrast for $\tilde r>>1$, 
$\Delta>>1$, fluctuations in $x_w$ is large; hence a DDW ensues.


\subsection{Steady density profiles in Model II}\label{density2}

Unlike Model IA and Model IB, sites in $\cal S$ of Model II can accommodate maximum one particle per site, i.e., exclusion is imposed on 
$\cal S$ as well. 
Nonetheless, the current $J_{\cal S}$ is still given by 
(\ref{diffcur}). Due to exclusion, densities $\gamma$ and $\delta$ cannot exceed $1$. In the steady state, when $J_{\cal S}=J_{\cal T}$, in order to have a finite current in TL, we must make $D\propto N$. As in Model IB,
we define  $D=Nd$, so that 
the current in $\cal S$ 
is 
written 
as 
\begin{equation}
 J_{\cal S}=(\gamma-\delta)d/r,
 \label{sepcur1}
\end{equation} 
as in Model IB.
In NESS, the particle current is constant throughout the system. Considering this at the two junctions of $\cal T$ and $\cal S$, we find the following two conditions:
\begin{eqnarray}
 \delta D (1-\alpha) &=& \alpha (1-\alpha),\\
 \gamma(1-\gamma)&=& (1-\beta) (1-\gamma).
 \end{eqnarray}
The first condition is common with Model IA and Model IB; the second one appears exclusively here. The latter one - unlike Model IA and IB - fixes $\beta$, i.e., relates it with $\gamma$. Thus $\beta$ is no longer a free parameter. Using these 
conditions 
one can arrive at the following relations:
\begin{equation}
 \alpha=D\delta,    \gamma=(1-\beta),
 \label{sepden}
\end{equation}
as found in Ref.~\cite{hinsch}. Thus, $\delta=\alpha/D$ approaches zero in TL 
since $D$ scales with $N$ and $\alpha\sim {\cal O}(1)$. In the main part of the paper, we primarily
concern 
ourselves with the nature of domain walls. Analysis of the other 
phases including the phase diagrams are made available in the Appendix.

For a domain wall, we set $\alpha=\beta$. 
This implies an active part current in the bulk
$J_{\cal T}=\alpha(1-\alpha)$. Hence Eq.~(\ref{sepcur1}) can be rewritten as 
\begin{equation}
 \delta=(1-\alpha)(1-\frac{r\alpha}{d})\rightarrow 0.
\end{equation}
This then gives $\alpha=\frac{d}{r}$. For a domain wall at $x_w$ in $\cal T$,  
$\rho(x)$ is given by Eq. (\ref{tasep-theta}); the steady density in $\cal S$ is still given by Eq. (\ref{sepdengen}).
Now from particle number 
conservation,
\begin{equation}
 n_p(1+r)=\int_{0}^{x_w}\alpha dx+\int_{x_w}^{1}(1-\beta)dx+\int_{0}^{r}
 [\delta+(\gamma-\delta)\tilde x/r]dx
\end{equation}
Solving these in the thermodynamic limit the DW position is given by
\begin{equation}
 x_w=\frac{2n_pr(1+r)-(r-d)(2+r)}{4d-2r}.
 \label{dw}
\end{equation}
MF solution (\ref{dw}) yields the {\em mean position} $x_w$ of the domain wall that is 
parametrised by $d,r,n_p$. This matches with the result in Ref.~\cite{hinsch} for 
$r=1$, for which an LDW is obtained, as expected. Surprisingly, for $r\gg 1$ the nature of the domain wall changes drastically, as revealed 
in Fig.~\ref{system1} 
below. Our MF analysis has been complemented by MCS. We have plotted $\rho(x)$ for $N=100$,
$r=5$ and $n_p=0.48$ 
for different diffusion coefficient $d$ in Fig. (\ref{system1}) (top). It is evident from the plot that for small $r$,  the domain wall is pinned, i.e., an LDW is observed. We have also plotted the 
$\rho(x)$ for $N=100$ with $r=N$ and $r=N/2$ for $n_p=0.48$ 
for different diffusion coefficient $d$ in Fig. (\ref{system1}) (bottom). For large values of $r$ a moving domain wall or DDW is found.
\begin{figure}[!htbp]
 \includegraphics[width=15.5cm,height=8.0cm,angle=0]{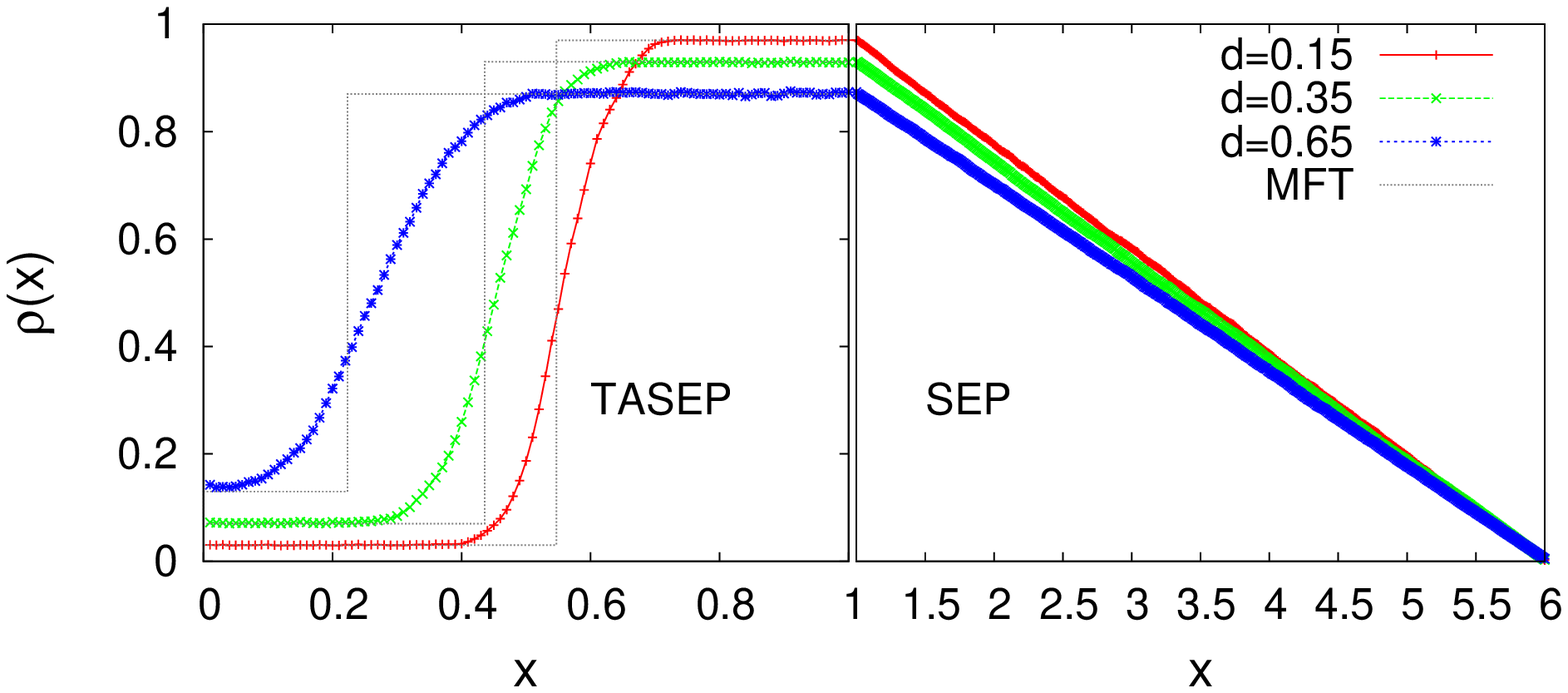}\\
  \includegraphics[width=15.5cm,height=8.0cm,angle=0]{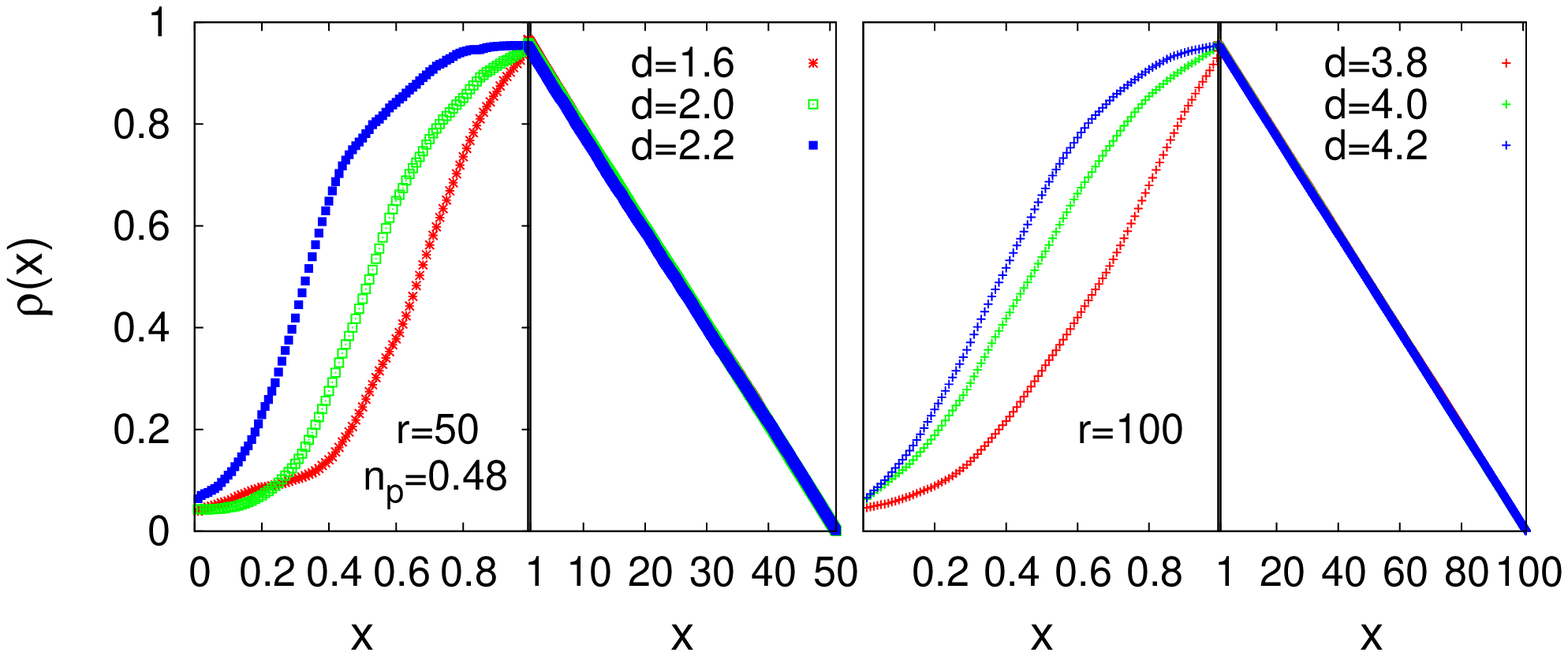}
  \caption{Gradual delocalisation of the DW in $\cal T$ in Model II. 
  Steady state plots of $\rho(x)$ versus $x$ for $N=100$, 
$r=5$ and $n_p=0.48$ (top); for $N=100$, 
$r=50$ and $n_p=0.48$ (bottom-left) and 
$r=100$ and $n_p=0.48$ (bottom-right) for different values of $d$. }
  \label{system1}
\end{figure}

\begin{figure}[!htbp]
 \includegraphics[width=10.5cm,height=7.0cm,angle=0]{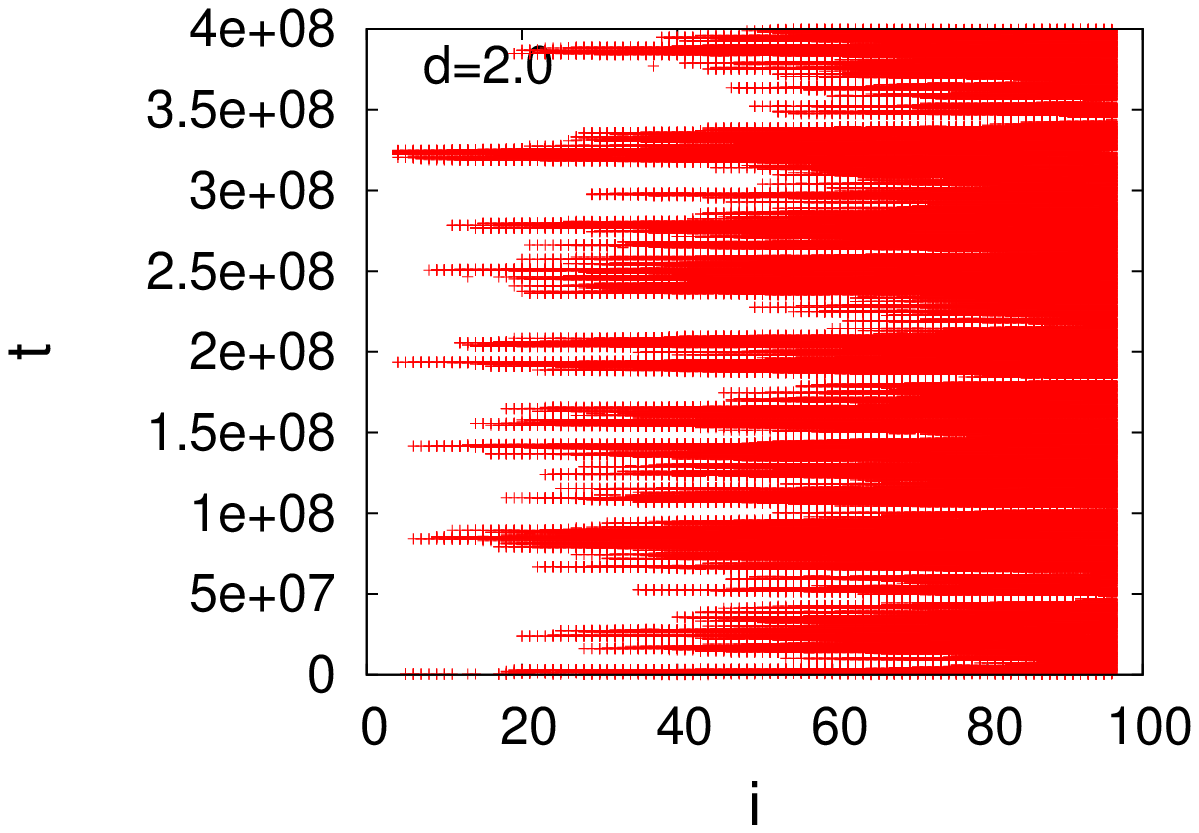}\\
  \includegraphics[width=10.5cm,height=7.0cm,angle=0]{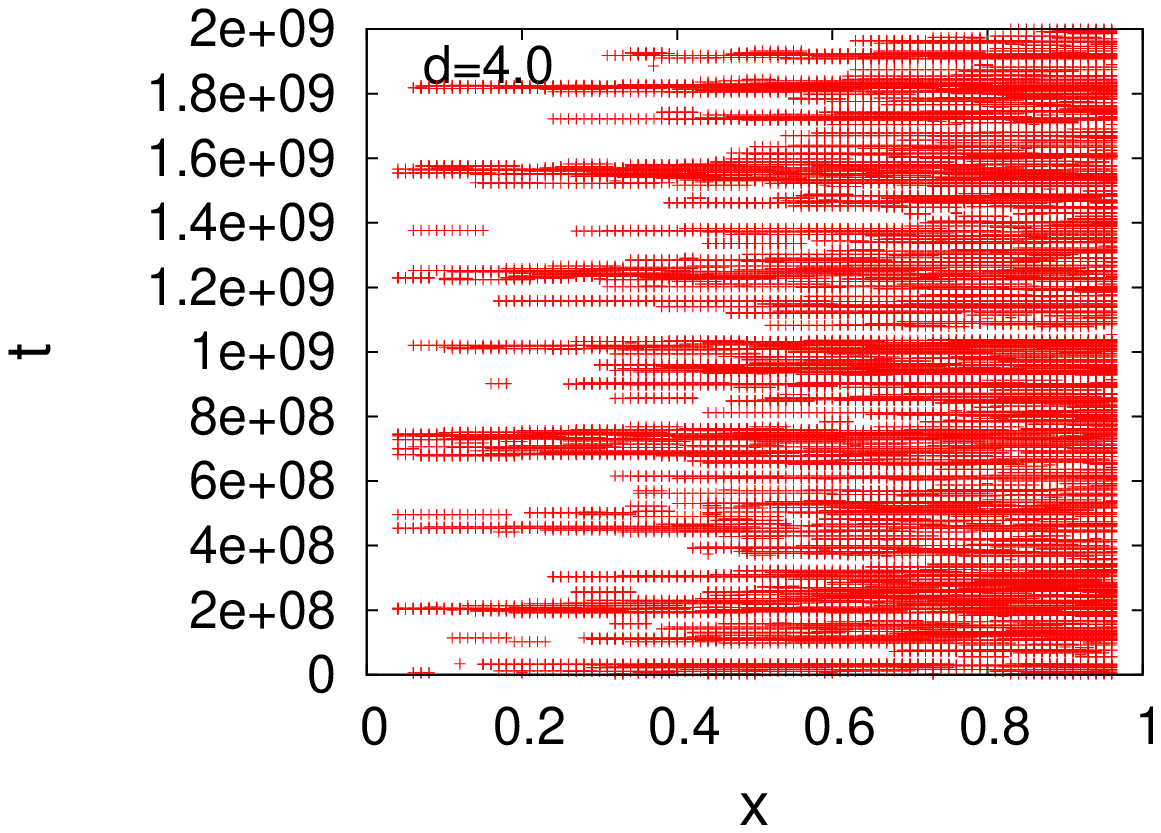}
  \caption{Kymograph for $N=100$ and $r=50$ for $d=2.0$ and $n_p=0.48$ (top) and $r=100$ for $d=4.4$ and $n_p=0.48$  (bottom). }
  \label{system_1}
\end{figure}

Thus, we notice that as $r$ rises for a fixed $d$ and $n_p$, the domain wall gradually delocalises 
and we eventually obtain a {\em single DDW} for $r\gg 1$. For $r=1$, we reproduce the results of Ref.~\cite{hinsch}.



Our explanation for the crossover from an LDW to a single DDW as $r$ rises to a large value runs exactly parallel to our discussions in Sec.~\ref{ldwddw}. As in Model IA and IB, the existence of a DDW that spans the whole of $\cal T$ implies particles number fluctuations $\Delta$ 
in $\cal T$ that should scale with $N$: $\Delta \sim {\cal O} (N)$. Conservation of total particle number implies equal and compensation particle number fluctuations to take place in $\cal S$. 
Assuming non-interacting particles, fluctuations $\Delta$ in the total number 
of particles in $\cal S$ should be ${\cal O} \sqrt{rN}$: $\Delta\sim \sqrt{rN}$. 
Clearly, in the limit $r\sim {\cal O}(1)$, $\Delta<< {\cal O} (N)$, fluctuations in $x_w$ is 
negligible; hence an LDW is observed. In contrast for $r\simeq N$, 
$\Delta\sim {\cal O} (N)$, i.e., the fluctuations in $x_w$ are large; hence a DDW ensues. 
The assumption of non-interacting particles in $\cal S$, however, remains somewhat questionable, since exclusion certainly introduces a hardcore repulsive interaction among the particles in $\cal S$. 


\section{Conclusions}\label{summary}

We have studied the steady states in periodic heterogeneous exclusion processes which is 
composed of one TASEP ($\cal T$) and one diffusive ($\cal S$) segment. We mainly focus on the nature of the domain wall in the system, which is usually known to be an LDW in similar closed heterogeneous TASEPs. We heuristically argue that a sufficiently large particle content in $\cal S$ can actually delocalise the LDW to a DDW. In order to establish the generality of our argument, we have 
constructed  three different choices for the dynamics in $\cal S$, which 
differ in the presence or absence of exclusion in $\cal S$, or finite or 
diverging diffusivity $D$ in the thermodynamic limit. The size of $\cal S$ relative to $\cal T$ 
is given by a model parameter $r$. Our MCS studies reveal a generic depinning 
of an LDW yielding a single DDW for sufficiently large $N_s$, the average total particle number in $\cal S$ in the steady states, which we justify by considering the fluctuations. This feature is shared by all three models we studied. To our 
knowledge, this is the first instance of a single DDW in a periodic 
exclusion process with strict particle number conservation. Not unexpectedly, 
this is not borne out by our MFT. We then argue that this depinning may be 
explained in terms of the particle number fluctuations in $\cal S$ that 
increases monotonically with $r$ and also with particle density $n_p$. This mechanism is shown to be quite robust as 
all the three models studied here display similar behaviour. In fact, we expect this mechanism to be generic - a pinned shock in any closed heterogeneous TASEP is expected to get depinned by this mechanism. For instance, in the well-known models for TASEPs with finite 
resources~\cite{tsep-zia}, where reservoirs do not have any internal dynamics of their own, a single LDW in the TASEP 
is expected to get delocalised by the mechanism illustrated here.  We thus close by remarking that we expect our studies to be potentially relevant to protein synthesis in a cell. In live eukaryotic cells, the supply of ribosomes (modeled by particles in
TASEP) is finite. For low resources, an LDW may be observed, where as for larger resources a DDW should follow. Since a DDW implies larger fluctuations not just in {\cal S} but also in {\cal T} (due to the overall particle number conservation), a DDW can be experimentally 
detected in standard ribosome density mapping experiments~\cite{ribo-profile}. We expect our studies will provide further impetus to future studies along these directions.

\section{Acknowledgement}
One of the authors  (A.B.) thanks the Alexander von Humboldt Stiftung, Germany 
for partial 
financial support 
through the Research Group Linkage Programme (2016).

\appendix

\begin{figure}[!htbp]
 \includegraphics[width=14.0cm,height=7.0cm,angle=0]{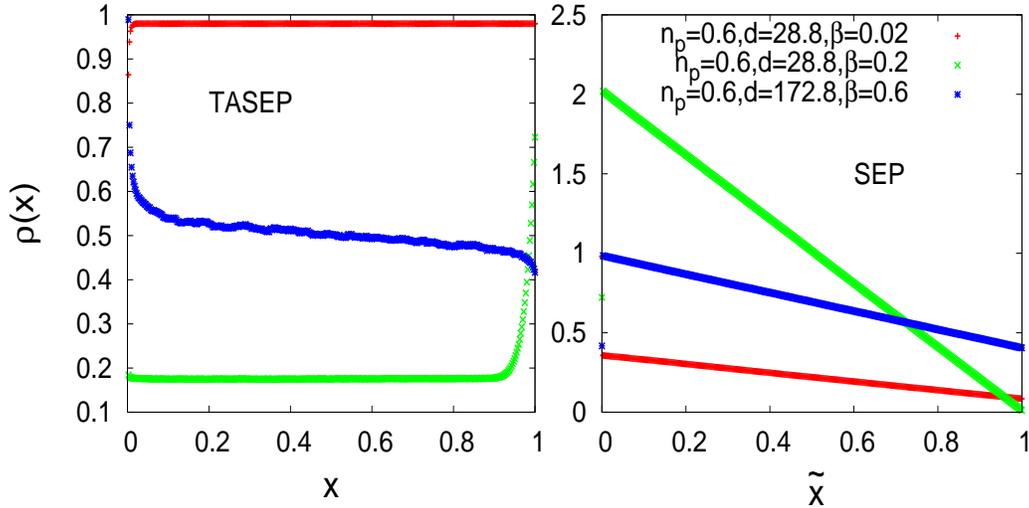}
  \caption{Plot of density profiles of Model IA for $N=400$, $r=1$ for HD, LD and MC phase. 
  The first half represents density profile in $\cal T$ and the second half represents density profile in $\cal S$.}
  \label{hd_ld_mc3}
  \end{figure}
  \begin{figure}[!htbp]
 \includegraphics[width=14.0cm,height=7.0cm,angle=0]{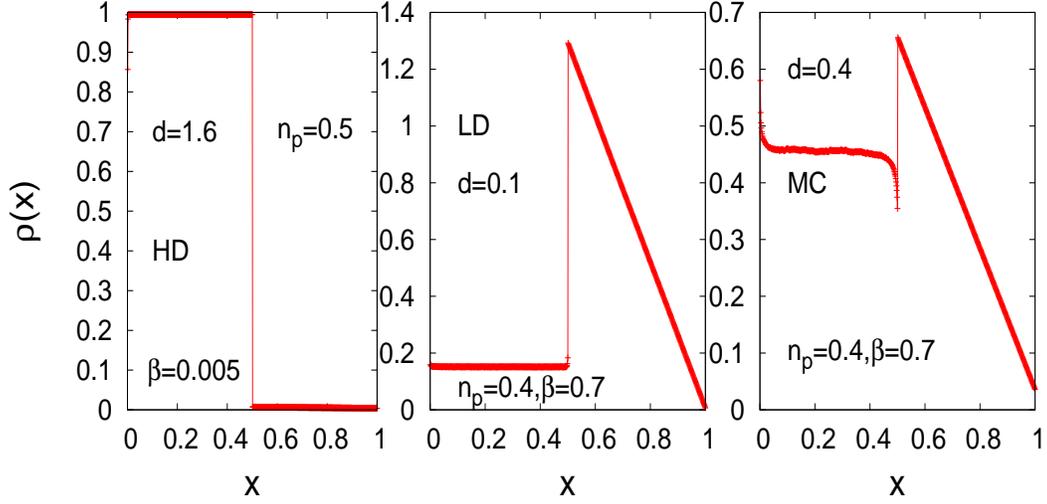}
  \caption{Plot of density profiles of Model IB for $N=400$, $r=1$ for HD, LD and MC phase.}
  \label{hd_ld_mc2}
\end{figure}

\begin{figure}[!htbp]
 \includegraphics[width=14.0cm,height=7.0cm,angle=0]{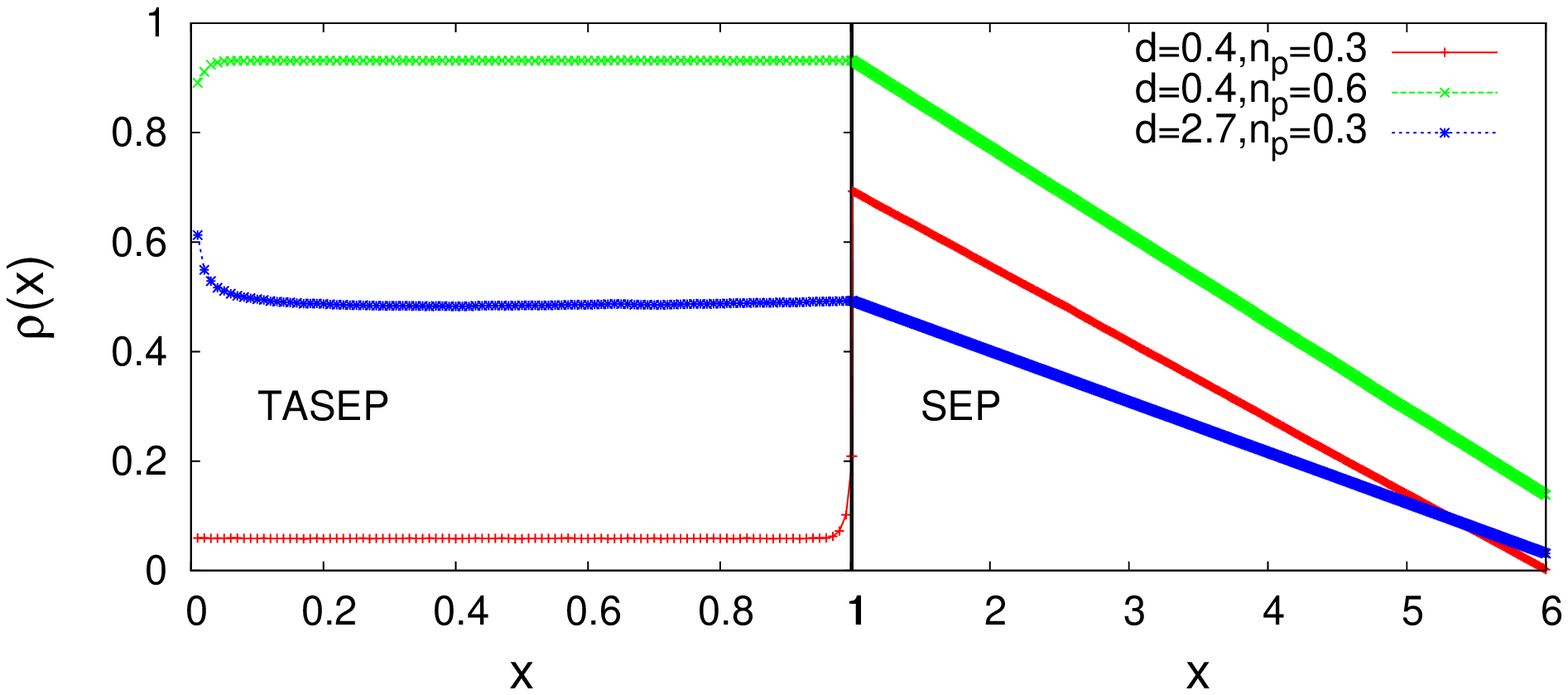}
  \caption{Plot of density profiles in $\cal T$ of Model II for $N=100$, $r=5$ for HD, LD and MC phase. }
  \label{hd_ld_mc1}
\end{figure}

\section{Domain walls in Model IA}
\label{dw1a}

We now provide the detailed derivation for the domain wall position $x_w$ in Model IA. Since multiple occupancy in the diffusive channel 
is allowed, $\gamma$, $\delta$ are unrestricted and can assume any value. Using (\ref{diffcur}) for $J_{\cal S}$ and 
equality of $J_{\cal S}$ and $J_{\cal T}$ at the two junctions of $\cal S$ and $\cal T$ implies
\begin{eqnarray}
 \alpha(1-\alpha)&=&(\gamma-\delta)D/Nr,\nonumber \\
 \label{apd2}
 D\delta(1-\alpha)&=&\alpha(1-\alpha)
\end{eqnarray}
Hence $\delta=\frac{\alpha}{D}\sim O(1)$.
Since particles from $\cal T$ are free to exit at a rate $\beta$ independent of  $\gamma$ at the first site of $\cal S$, there is no connection 
between $\gamma$ 
and $\beta$.
Using $\delta=\frac{\alpha}{D}$ we have from Eq.~(\ref{apd2}),
\begin{equation}
 \gamma=[\alpha(1-\alpha)+\frac{\alpha}{Nr}]\frac{Nr}{D}.
\end{equation}
From particle number conservation one can write,
\begin{equation}
 N_{p}=\int_{0}^{1}\rho(x)Ndx+\int_{0}^{r}\tilde{\rho}(\tilde x)Nd\tilde x
\end{equation}
Let $x_w$ be the position of the DW. Then density distribution in the TASEP channel 
$\rho(x)=\alpha+\Theta(x-x_w)(1-\alpha-\beta)$ and that in the diffusive channel $\tilde{\rho}(\tilde x)=\delta+(\gamma-\delta)\tilde x/r$. 
For a domain wall to exist $\alpha=\beta$. These 
allow us to write,
\begin{eqnarray}
 n_p(1+r)&=&\int_{0}^{x_w}\alpha dx+\int_{x_w}^{1}(1-\alpha)dx+\int_{0}^{r}[\delta+(\gamma-\delta)x/r]dx\nonumber\\
        &=& x_w(2\alpha-1)+(1-\alpha)+r\delta/2+r\gamma/2\nonumber\\
        &=& x_w(2\alpha-1)+(1-\alpha)+\frac{Nr^2}{2D}[\alpha(1-\alpha)+\frac{\alpha}{Nr}]+\frac{r\alpha}{2D}.
\end{eqnarray}  
Hence,
\begin{equation}
 x_w=\frac{n_p(1+r)-(1-\alpha)-\frac{Nr^2}{2D}[\alpha(1-\alpha)]-\frac{r\alpha}{D}}{(2\alpha-1)}
 \label{scdw4}
\end{equation}
gives the position of the domain wall in MFT.

\subsection{Phase diagram of Model IA}
\label{pd1a}

As mentioned earlier, the phase space of Model IA is spanned by three parameters - $n_p, D$ and $\beta$. As explained above, 
for $\beta <1/2$, LD and HD are the only possible phases, with the possibilities of an LD-HD coexistence phase in the form of a domain wall.  
The phase boundaries between the LD, LD-HD and HD phases may be obtained as follows.

The boundary between the LD and the LD-HD coexistence region may be obtained by setting $x_w=1$ in   Eq.~(\ref{scdw4}). This gives 
\begin{equation}
 D=\frac{Nr^{2}\beta(1-\beta)+2r\beta}{2[n_p(1+r)-\beta]}.
\end{equation}
Similarly, the phase boundary between HD and LD-HD phase is 
obtained by setting $x_w=0$ in Eq.~(\ref{scdw4}), giving
\begin{equation}
 D=\frac{Nr^{2}\beta(1-\beta)+2r\beta}{2[n_p(1+r)-(1-\beta)]}.
\end{equation}
 Figure~\ref{phase1b} shows the MCS and MFT phase diagram of Model IA in the 
($n_p-D$) plane for $r=1$ and $\beta=0.1$.



For $\beta>1/2$, it has been argued that only LD and MC phases are possible with no HD phase. The phase boundary between the LD and MC 
phases in the $n_p-D$ plane for a given $\beta$ may be obtained as follows.

\begin{figure}[!htbp]
 \includegraphics[width=10.5cm,height=7.0cm,angle=0]{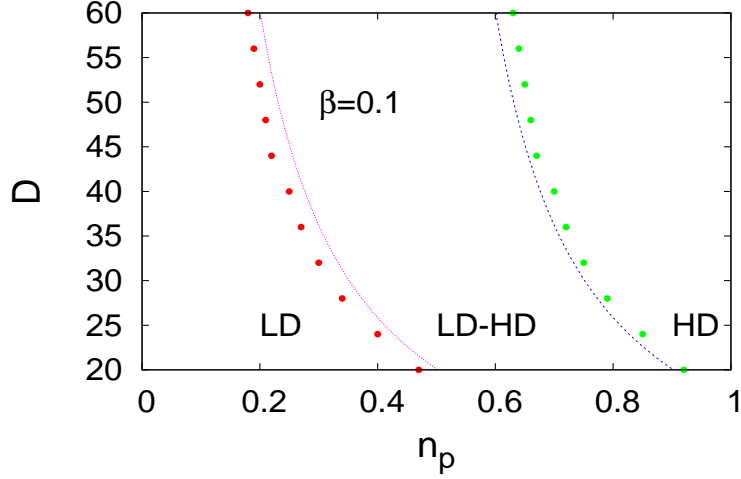}
  \caption{Solid lines represent the phase diagram obtained using MFT and the 
circles represent the same obtained using MCS for Model IA for $\beta=0.1$.}
  \label{phase1a1}
\end{figure}

\begin{figure}[!htbp]
 \includegraphics[width=10.5cm,height=7.0cm,angle=0]{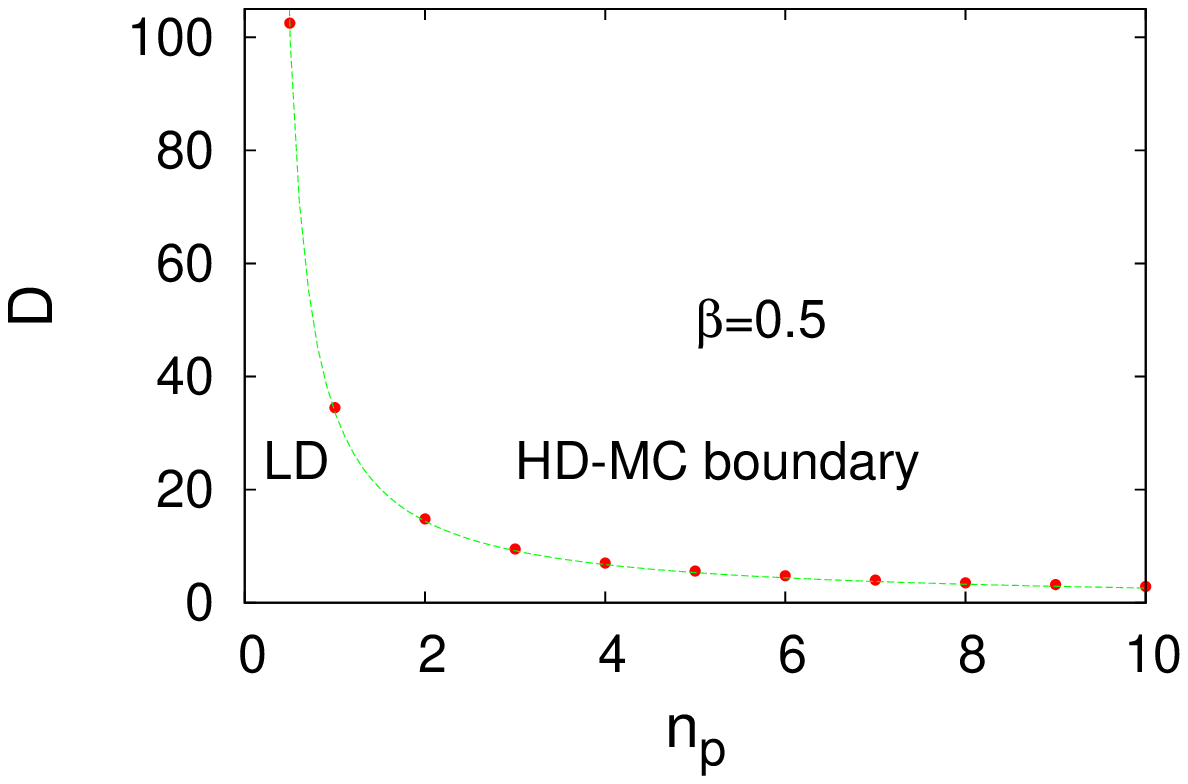}
  \caption{Solid lines represent the phase diagram obtained using MFT and the 
circles represent the same obtained using MCS for Model IA for $\beta=0.5$.}
  \label{phase1a2}
\end{figure}

\begin{figure}[!htbp]
 \includegraphics[width=10.5cm,height=7.0cm,angle=0]{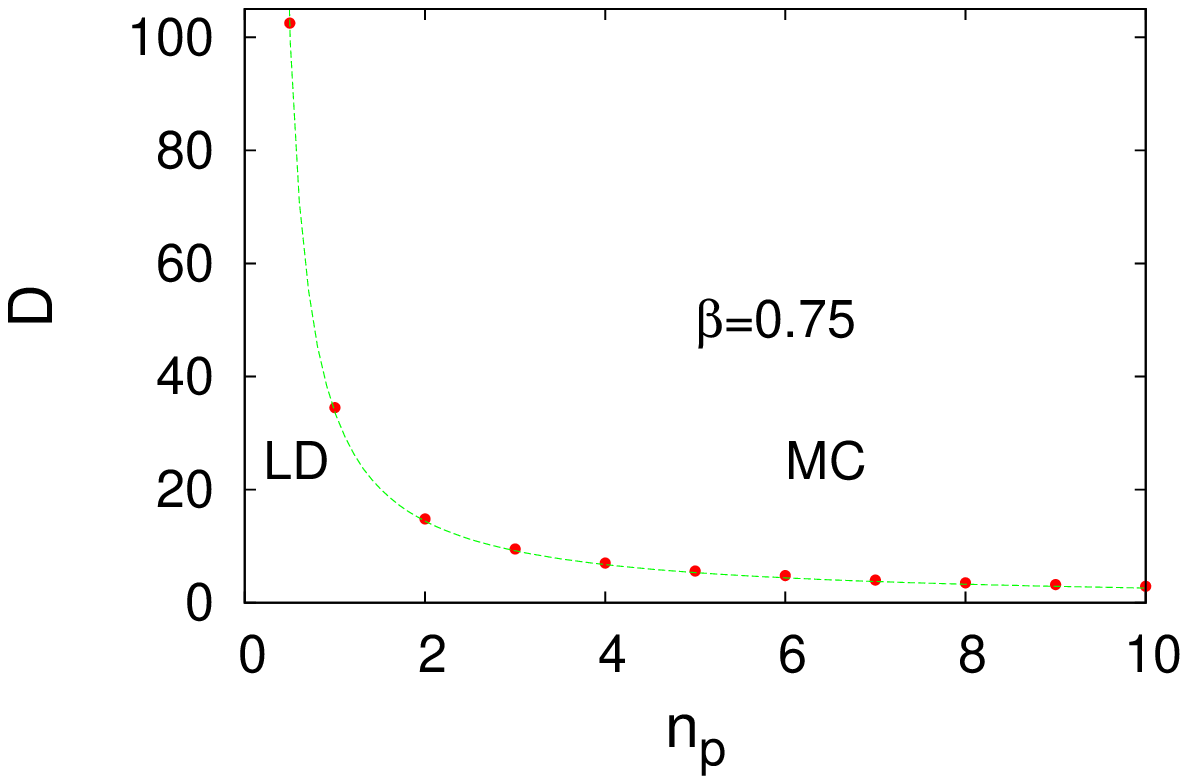}
  \caption{Solid lines represent the phase diagram obtained using MFT and the 
circles represent the same obtained using MCS for Model IA for $\beta=0.75$.}
  \label{phase1a3}
\end{figure}

In LD phase bulk density in $\cal T$ is $\rho(x)=\alpha$ and in  MC phase $\rho(x) =1/2$. In $\cal S$ is $\tilde\rho(\tilde x)=
\delta+(\gamma-\delta)\frac{\tilde x}{r}$. From particle number conservation
\begin{equation}
 n_p(1+r)=\alpha\int_{0}^{1}dx+\int_{0}^{r}[\delta+(\gamma-\delta)\tilde x/r]d\tilde x
 \label{PNC}
\end{equation}
Current continuity at the 
junction A gives $\delta D (1-\alpha)=\alpha(1-\alpha)$. Using this 
condition 
one can write, $\delta=\frac{\alpha}{D}$.
  Equality of $J_{\cal T}$ and $J_{S}$ in the NESS implies $\alpha(1-\alpha)=(\gamma-\delta)D/Nr.$
Hence, $\gamma=\alpha(1-\alpha)\frac{Nr}{D}+\frac{\alpha}{D}.$ Using these equations for $\gamma$ and $\delta$ one can obtain from 
Eq.~(\ref{PNC})
\begin{equation}
 n_p(1+r)=\alpha+\alpha\frac{r}{D}+\frac{\alpha(1-\alpha)}{D}\frac{Nr^2}{2}
 \label{bldmc}
\end{equation}
For $\beta>1/2$ as long as $\alpha<1/2$ we satisfy the condition for LD phase. As $\alpha$ exceeds $\frac{1}{2}$, we satisfy the condition 
for MC phase.
Thus the phase boundary in the $n_p-D$ plane between the LD and MC phases is obtained by substituting $\alpha=\frac{1}{2}$ in Eq.~(\ref{bldmc}) as,
\begin{equation}
 n_p(1+r)=\frac{1}{2}+\frac{r}{2D}+\frac{r^2N}{8D}.
 \label{ldmc1}
\end{equation}
For $\beta =1/2$, when $\alpha <1/2$ LD phase ensues with a bulk density $\rho(x)=\alpha$, whereas for $\alpha >1/2$ in a given region of the phase space in the $n_p-D$ plane, the bulk density in $\cal T$ is $\rho =1/2$ with a {\em boundary layer} at $x=0$. This is reminiscent of 
the density profile in an open TASEP at the boundary of the MC and HD phases. See phase diagrams in Figs. (\ref{phase1a1}, \ref{phase1a2}, \ref{phase1a3}),
our MFT and MCS results agree well.


\section{Domain walls in Model IB}
\label{dw1b}
Now we provide the detailed derivation for the position of the DW $x_w$ in model IB. When $D$ scales with system size $N$ we set $D=dN$ 
with $d\sim {\cal O}(1)$.
Current continuity at the junction A gives $\delta D (1-\alpha)=\alpha(1-\alpha)$. Using this 
condition one can write, $\delta=\frac{\alpha}{D}\sim {\cal O}(1/N)$. Here $\delta\to 0$ in the TL which is different from model IA.

Here we impose the exit rate $\beta$ of TASEP from outside. Thus $\beta$ is a 
free parameter. 
In contrast entrance rate $\alpha$ is to be derived from the 
conditions of steady state. We focus on the formation of DW for which $\alpha = 
\beta$.
Equality of $J_{\cal T}$ and $J_{S}$ in the NESS implies
\begin{equation}
 \alpha(1-\alpha)=(\gamma-\delta)D/Nr.
 \label{sepeqtasep}
\end{equation}
Using $\delta\to 0$ in the TL we have from Eq. (\ref{sepeqtasep}),
\begin{equation}
 \gamma=\frac{Nr}{D}[\alpha(1-\alpha)]. 
 \label{gamma}
\end{equation}
For a DW to exist, one must have $\alpha =\beta < 1/2$.






From particle number conservation one can write,
\begin{equation}
 N_{p}=\int_{0}^{1}\rho(x)Ndx+\int_{0}^{r}\tilde{\rho}(x)Nd\tilde x
\end{equation}
Let $x_w$ be the position of the DW. Now approaching as in Model IA, we have
\begin{eqnarray}
 n_p(1+r)&=&\int_{0}^{x_w}\alpha dx+\int_{x_w}^{1}(1-\alpha)dx+\int_{0}^{r}[\delta+(\gamma-\delta)x/r]dx\nonumber\\
        &=& x_w(2\alpha-1)+(1-\alpha)+r\delta/2+r\gamma/2\nonumber\\
\end{eqnarray}  

Here in the TL, $\delta\to 0$. 
Using the expression of $\gamma$ from eq. (\ref{gamma}) the location $x_w$ of the domain wall is 
given by
\begin{equation}
 x_w=\frac{n_p(1+r)-(1-\alpha)-\frac{r^2}{2d}[\alpha(1-\alpha)]}{(2\alpha-1)}
 \label{scdw3}
\end{equation}


\subsection{Phase diagram of Model IB}
\label{pd1b}

As mentioned earlier, the phase space of Model IB is spanned by three parameters - $n_p, d=D/N$ and $\beta$. As explained above, 
for $\beta <1/2$, LD and HD are the only possible phases, with the possibilities of an LD-HD coexistence phase in the form of a domain wall.  
The phase boundaries between the LD, LD-HD and HD phases may be obtained as follows.

For $x_w\le0$ the DW leave the active 
part at the left junction resulting in HD phase having constant density 
$\rho(x)=1-\beta$, $x_w\ge1$ results in LD phase having constant density $\rho(x)=\alpha$ 
and LD-HD phase is characterised by the presence of a DW inside the system for 
$0<x_w<1$. 

Hence the boundary between the HD and LD-HD coexistence region may be obtained by setting $x_w=0$ in Eq. (\ref{scdw3}). This gives
\begin{equation}
 d=\frac{r^{2}\beta(1-\beta)}{2[n_p(1+r)-(1-\beta)]}.
\end{equation}
Similarly the boundary between LD and LD-HD coexistence region is obtained by setting $x_w=1$ in Eq. (\ref{scdw3}) which gives 
\begin{equation}
 d=\frac{r^{2}\beta(1-\beta)}{2[n_p(1+r)-\beta]}.
\end{equation}
 
 Our main results are summarised in the phase diagrams in the 
($n_p-d$) plane in Fig.~\ref{phase1a} for $r=1$ and $\beta=0.2$. For $\beta>1/2$, only 
LD and MC phases are possible with no HD phase. The phase boundary between LD and MC phases in the $n_p$-$d$ plane for a given $\beta$ may 
be obtained as follows. 

\begin{figure}[!htbp]
 \includegraphics[width=10.5cm,height=7.0cm,angle=0]{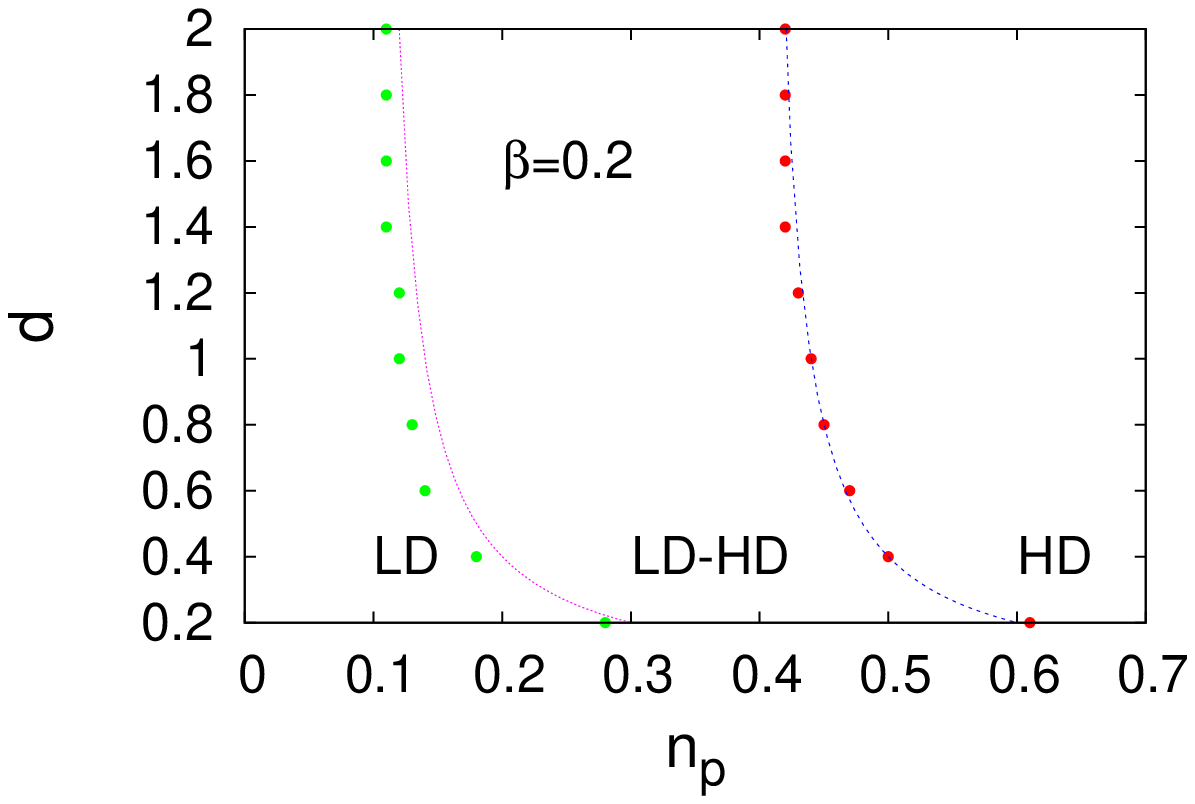}
  \caption{Solid lines represent the phase diagram obtained using MFT and the 
circles represent the same obtained using MCS for Model IB for $\beta=0.2$.}
  \label{phase1b1}
\end{figure}

\begin{figure}[!htbp]
 \includegraphics[width=10.5cm,height=7.0cm,angle=0]{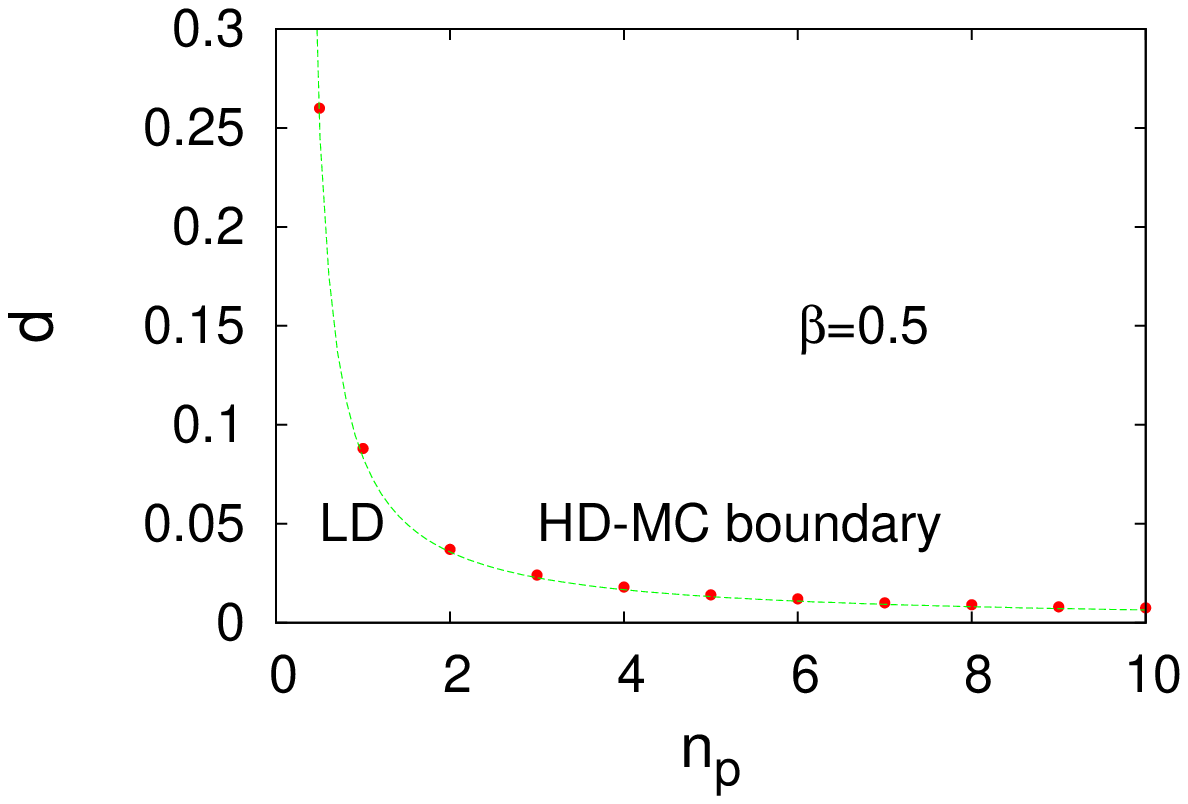}
  \caption{Solid lines represent the phase diagram obtained using MFT and the 
circles represent the same obtained using MCS for Model IB for $\beta=0.5$.}
  \label{phase1b2}
\end{figure}

\begin{figure}[!htbp]
 \includegraphics[width=10.5cm,height=7.0cm,angle=0]{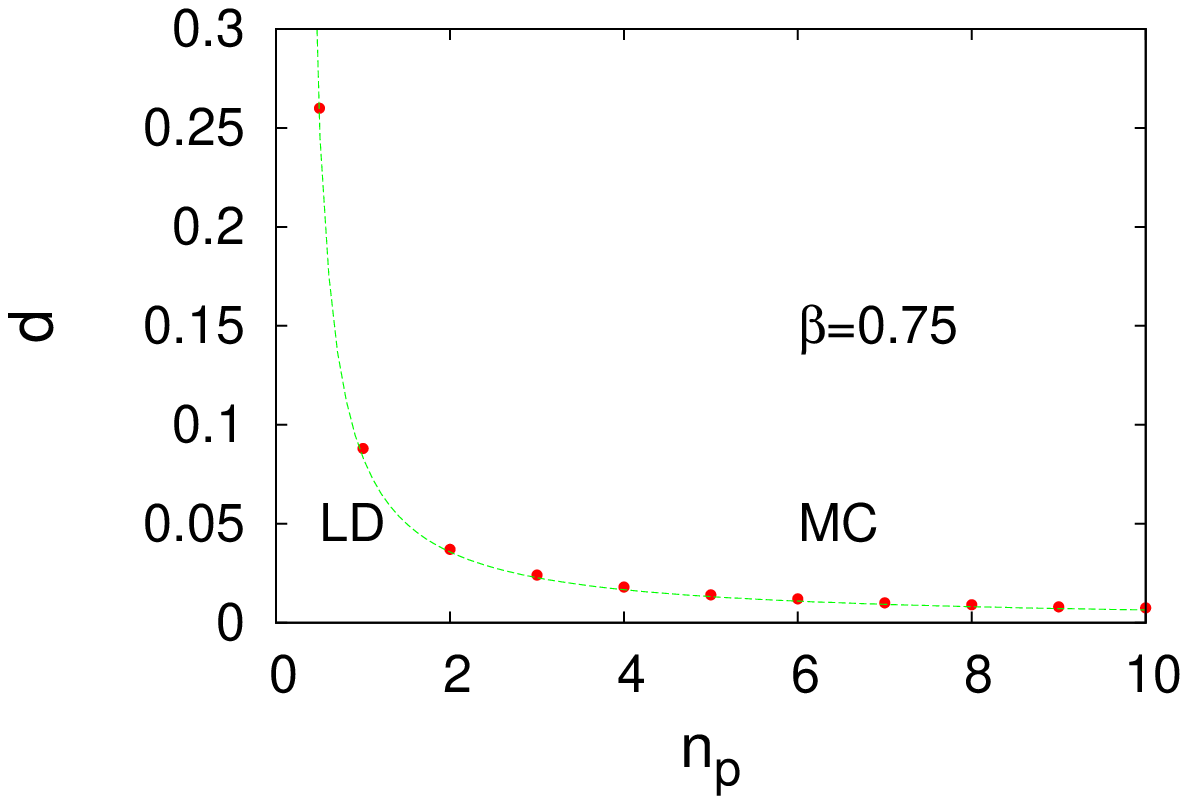}
  \caption{Solid lines represent the phase diagram obtained using MFT and the 
circles represent the same obtained using MCS for Model IB for $\beta=0.75$.}
  \label{phase1b3}
\end{figure}

In the LD phase bulk density in $\cal T$ is $\rho(x)=\alpha$ and in MC phase $\rho(x)=1/2$. In $\cal S$ is $\tilde\rho(\tilde x)=
\delta+(\gamma-\delta)\frac{\tilde x}{r}$. From PNC 
\begin{equation}
 n_p(1+r)=\alpha\int_{0}^{1}dx+\int_{0}^{r}[\delta+(\gamma-\delta)\tilde x/r]d\tilde x
 \label{PNC1b}
\end{equation}
Using the expressions of $\gamma$ and $\delta$ for Model IB in Eq. (\ref{PNC1b}) one can obtain 
\begin{equation}
 n_p(1+r)=\alpha+\frac{\alpha(1-\alpha)}{D}\frac{Nr^2}{2}
 \label{bldmc1b}
\end{equation}
For $\beta>1/2$ as long as $\alpha<1/2$ we satisfy the condition for LD phase. As $\alpha$ exceeds $\frac{1}{2}$, we satisfy the condition 
for MC phase.
Thus the phase boundary in the $n_p-D$ plane between the LD and MC phases is obtained by substituting $\alpha=\frac{1}{2}$ in Eq.~(\ref{bldmc1b}) as,
\begin{equation}
 n_p(1+r)=\frac{1}{2}+\frac{r^2N}{8D}.
 \label{ldmc1b}
\end{equation}
For $\beta =1/2$, when $\alpha <1/2$ LD phase ensues with a bulk density $\rho(x)=\alpha$, whereas for $\alpha >1/2$ 
in a given region of the phase space in the $n_p-D$ plane, the bulk density in $\cal T$ is $\rho =1/2$ with a {\em boundary layer} at $x=0$. This is reminiscent of 
the density profile in an open TASEP at the boundary of the MC and HD phases. See phase diagrams in Figs 
(\ref{phase1b1}, \ref{phase1b2}, \ref{phase1b3}); our MFT and MCS results agree well.

\section{Domain walls in Model II}
\label{dw2}

From particle number conservation one can write,
\begin{equation}
 N_{p}=\int_{0}^{1}\rho(x)Ndx+\int_{0}^{r}\tilde{\rho}(x)Ndx
\end{equation}
Density distribution in the TASEP channel $\rho(x)=\alpha+\Theta(x-x_w)(1-\alpha-\beta)$ and that in the diffusive channel 
$\tilde{\rho}(x)=\delta+(\gamma-\delta)x/r$.
This implies,
\begin{eqnarray}
 n_p(1+r)&=&\int_{0}^{x_w}\alpha dx+\int_{x_w}^{1}(1-\beta)dx+\int_{0}^{r}[\delta+(\gamma-\delta)x/r]dx\nonumber\\
         &=&\alpha x_w+(1-\beta)(1-x_w)+\frac{r\gamma}{2}
 \label{ap1}        
\end{eqnarray}
Since $\delta\sim O(1/N)$ in the TL, $\delta\to 0$ and $\alpha\to \frac{d}{r}$. Also from Eq.(\ref{sepden}) $\gamma=1-\beta$. Now 
from the condition of DW 
$\alpha=\beta$. Using these conditions we have from Eq.(\ref{ap1}),
\begin{equation}
 x_w=\frac{2r(1+r)n_p-(r-d)(2+r)}{4d-2r}
\label {dwm2}
\end{equation}




\subsection{Phase diagrams for Model II}

For different values of $r$ we can get phase diagram in ($n_p$, $d$) space consisting of 
different regimes depending on the position of DW. For $x_w\le0$ the DW leave 
the active 
part at the left junction resulting in HD phase having constant density 
$\rho(x)=1-\beta$, $x_w\ge1$ results in LD phase having constant density 
$\rho(x)=\alpha$ 
and LD-HD phase is characterised by a DW localised inside the system for 
$0<x_w<1$. From Eq. (\ref{dwm2}) the phase boundary between HD and LD-HD coexistence is 
given by 
\begin{equation}
 d=r-\frac{n_p2r(1+r)}{2+r}.
\end{equation}
Phase boundary between LD and LD-HD phase is obtained as,
\begin{equation}
 d=\frac{2n_pr(1+r)-r^2}{2-r}
\end{equation}
 Similar to ordinary TASEP the ring system also exhibits maximal current (MC) 
phase which is characterised by maximam active part current $J_{MC}=1/4$ and 
constant 
 density $\rho=1/2$ and is 
 obtained for $\alpha$, $\beta>1/2$. Hence, $N/2$ particles have to be present 
in the active part. Equality of the active part current and passive part current 
allows 
 one to write $\gamma=\delta+\frac{r}{4d}.$
 
 Particle number conservation in MC phase yields,
 \begin{equation}
  (1+r)n_p=r(\delta+\frac{r}{8d})+\frac{1}{2}.
  \label{MC}
 \end{equation}
 
 The constraints $\alpha$, $\beta>1/2$ on active part impose constraints on the 
passive part by the use of Eqs. (\ref{sepcur1}) and (\ref{sepden}) as 
 \begin{equation}
  \delta>\frac{1}{2dN}
  \label{gr}
 \end{equation}
 and
 \begin{equation}
  \delta<\frac{1}{2}-\frac{r}{4d}.
  \label{lt}
 \end{equation}
 Implementing Eq. (\ref{gr}) in Eq. (\ref{MC}) one can get phase boundary 
between LD and MC phase in the TL as,
 \begin{equation}
  d=\frac{r^2}{8(1+r)n_p-4}.
 \end{equation}
 Similarly the phase boundary between HD and MC phase can be obtained using Eq. 
(\ref{lt}) in Eq. (\ref{MC}) as,
 \begin{equation}
  d=\frac{r^2}{4(1+r)(1-2n_p)}.
 \end{equation}

\begin{figure}[!htbp]
 \includegraphics[width=10.5cm,height=7.0cm,angle=0]{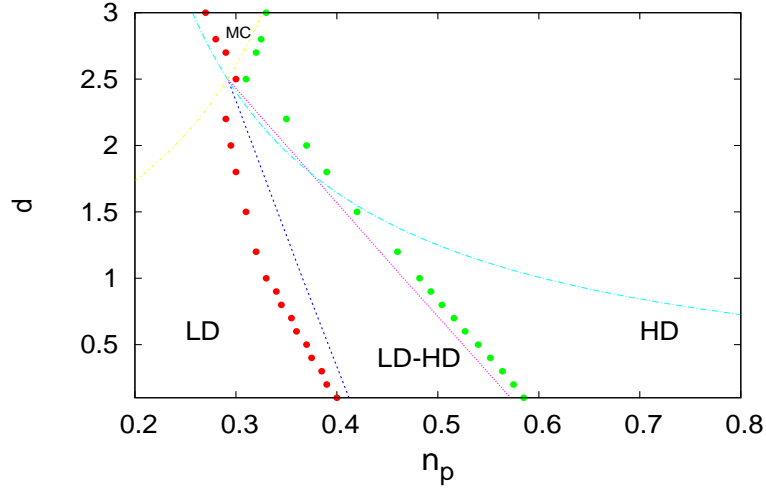}
  \caption{Solid lines represent the phase diagram obtained using MFT and the 
circles represent the same obtained using MCS for Model II for $r=5$. The phase diagram 
exhibits four
  different phases.}
  \label{phase2a}
\end{figure}

\begin{figure}[!htbp]
 \includegraphics[width=14.0cm,height=7.0cm,angle=0]{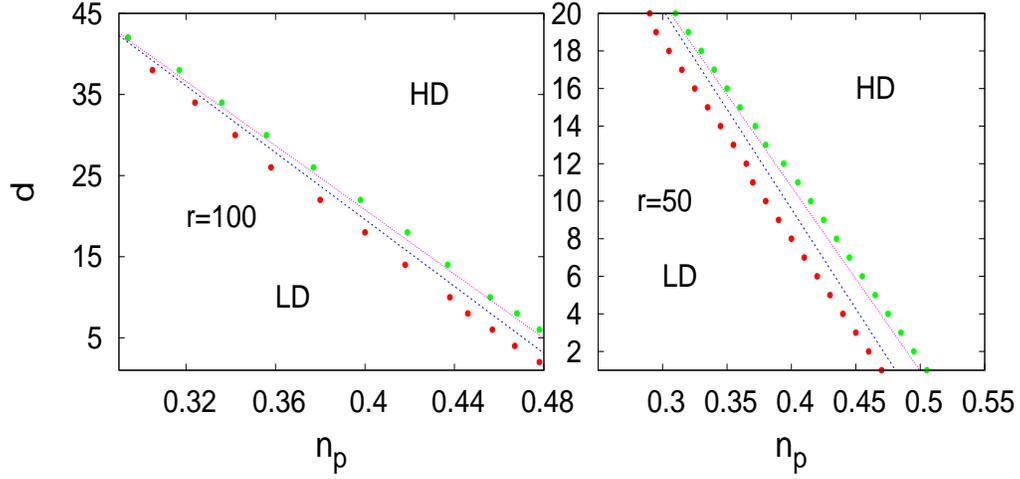}
  \caption{Solid lines represent the phase diagram obtained using MFT and the 
circles represent the same obtained using MCS for Model II. Left panel represents phase diagram for $r=100$ and right panel 
represents phase diagram for $r=50$.}
  \label{phase2b}
\end{figure}

Our main results are summarised in the phase diagrams in the 
($n_p-d$) plane in Fig.\ref{phase2a} and \ref{phase2b} for different values of $r$.

\begin{figure}[!htbp]
 \includegraphics[width=10.5cm,height=7.0cm,angle=0]{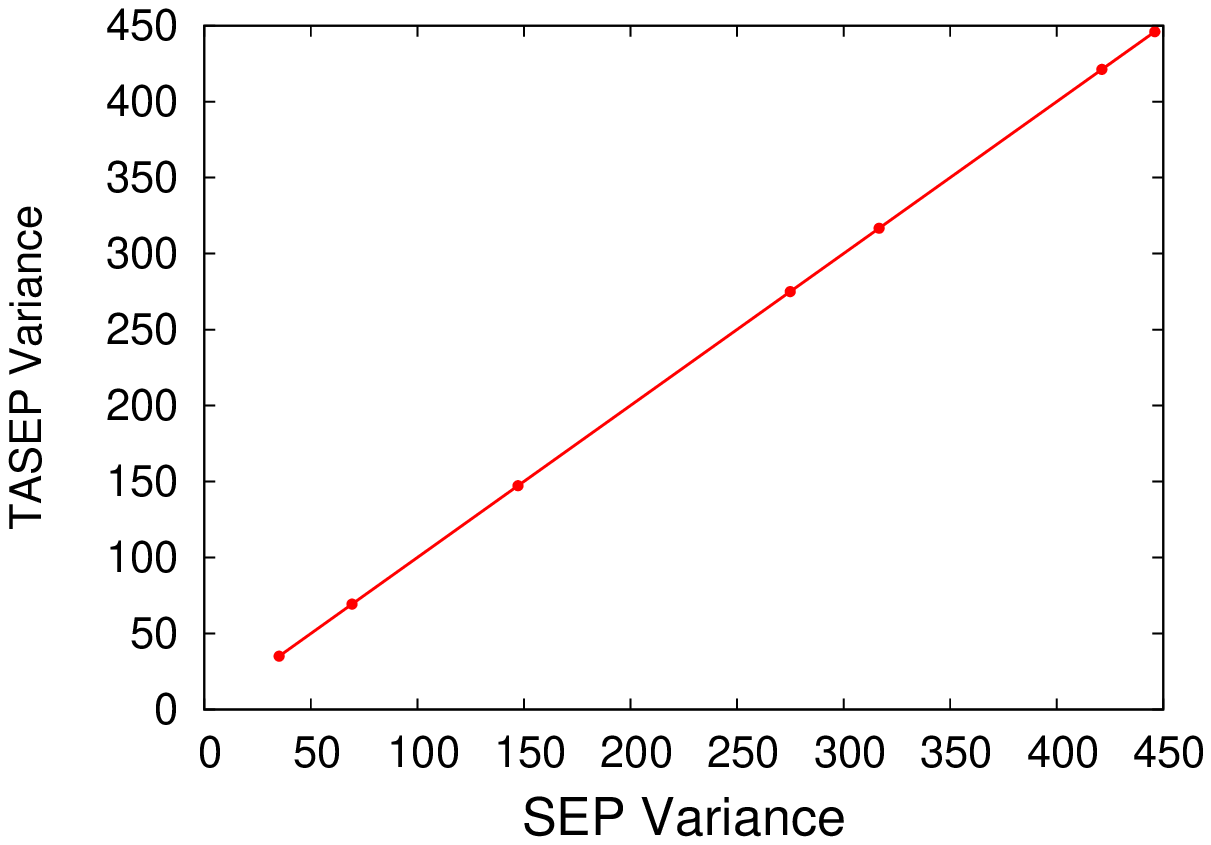}
  \caption{Plot of variance of TASEP vs SEP for Model 1A.}
  \label{var1a}
\end{figure}

\begin{figure}[!htbp]
 \includegraphics[width=10.5cm,height=7.0cm,angle=0]{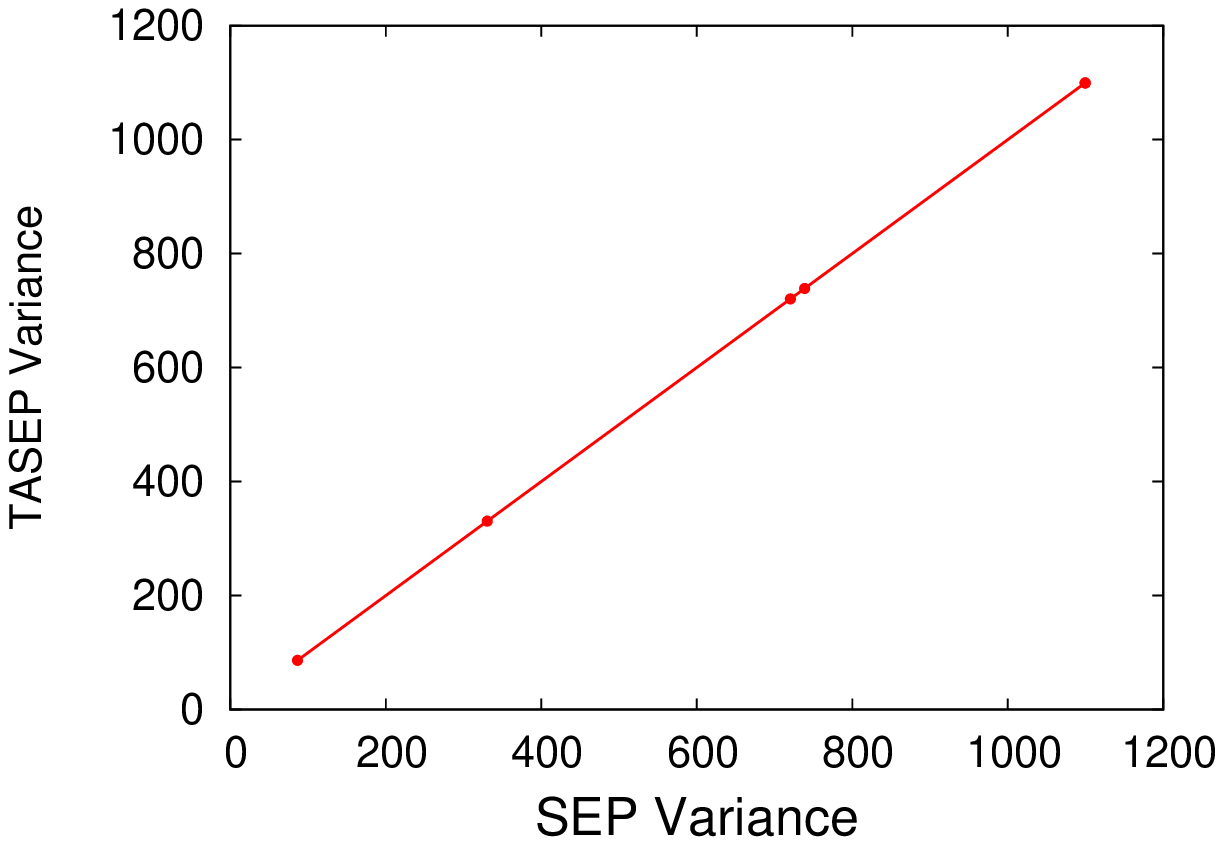}
  \caption{Plot of variance of TASEP vs SEP for Model 1B.}
  \label{var1b}
\end{figure}

\begin{figure}[!htbp]
 \includegraphics[width=10.5cm,height=7.0cm,angle=0]{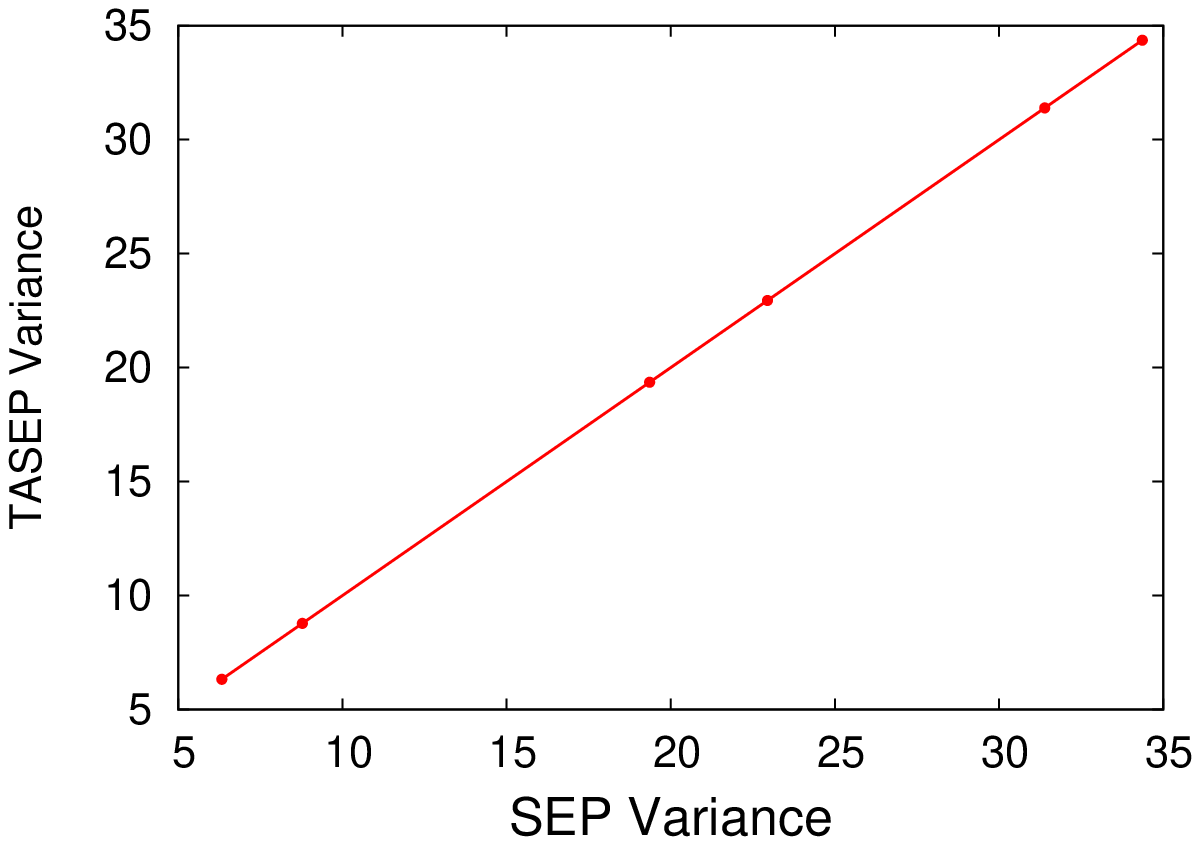}
  \caption{Plot of variance of TASEP vs SEP for Model 2.}
  \label{variance2}
\end{figure}


\begin{thebibliography}{99}
\bibitem{foot11} The two choices of $D\sim {\cal O}(1)$ and $D\sim {\cal O} (N)$ are two just two possible choices (among many others) 
that allow for finite steady state currents. These suffice for our purposes here.
\end{thebibliography}
\end{document}